\documentclass[%
 reprint,
 amsmath,amssymb,
 aps,
]{revtex4-2}

\usepackage{graphicx}
\usepackage{dcolumn}
\usepackage{bm}
\usepackage{color}
\usepackage{aas_macros}

\usepackage{lineno}

\begin{document}

\preprint{APS/123-QED}

\title{
A new understanding of nuclei spectra properties observed by AMS-02 experiment
}

\author{Xu-Lin Dong}
 \affiliation{ Hebei Normal University, Shijiazhuang, 050024, Hebei, China
 }
 \affiliation{%
Institute of High Energy Physics, Chinese Academy of Sciences,  Beijing, 100049, Beijing, China
}

\author{Yu-Hua Yao}%
\email{yaoyh@cqu.edu.cn}
\affiliation{%
Chongqing University, Chongqing, 401331, Chongqing, China
}%
\affiliation{%
 Institute of High Energy Physics, Chinese Academy of Sciences, Beijing, 100049, Beijing, China
}

\author{Yi-Qing Guo}
\email{guoyq@ihep.ac.cn}
\affiliation{%
 Institute of High Energy Physics, Chinese Academy of Sciences, Beijing, 100049, Beijing, China
}
\affiliation{
 University of Chinese Academy of Sciences, Beijing, 100049, Beijing, China
}

\author{Shu-Wang Cui}
\email{cuisw@hebtu.edu.cn}
\affiliation{%
Hebei Normal University, Shijiazhuang, 050024, Hebei, China
}

\date{\today}
\begin{abstract}
The AMS-02 experiment has observed new properties of primary cosmic rays (CRs) categorized into two groups: He-C-O-Fe and Ne-Mg-Si-S, which are independent of CR propagation. In this study, we investigate the unexpected properties of these nuclei using a spatial propagation model. All nuclei spectra are accurately reproduced and separated into primary and secondary contributions. Our findings include: 1. Primary CR spectra are identical. 2. Our calculations align with AMS-02 results for primary-dominated nuclei within a 10\% difference, but show significant discrepancies for the secondary-dominated nuclei. 3. The primary element abundance at around 200 GeV is presented and compared with previous solar and Galactic results. We hope that future DAMPE experiments can provide more experimental observational evidence to validate our model calculations.
\end{abstract}

\maketitle


\section{Introduction}
The discovery of cosmic rays (CR) has spanned over a hundred years, yet their origin, acceleration, and propagation remain ambiguous. Primary CR nuclei are generally believed to be accelerated by astrophysical sources, such as supernova remnants \citep{1934PNAS...20..254B}, or pulsars \citep{2022JCAP...10..081B}. Secondary CRs are produced through spallation reactions taking place at the production site or in the interstellar medium on their way to Earth. The CR spectrum and chemical composition, for both primaries and secondaries, provide the most important clues to cosmic-ray origin and propagation \citep{2015ARA&A..53..199G,2019ApJS..245...30L}. They can trace effects within CR sources by probing the average CR residence time and gas density inside accelerating sites, and can also trace a change in the diffusion coefficient between the Galactic disc and halo. Elemental abundance can also give essential clues to the acceleration sites and timescales.

The latest generation of experiments is currently delving into the intricate details of CR phenomenology. An observed and confirmed hardening from the uniform power-law of the CR spectrum for all elements around a rigidity of a few hundred GeV has sparked significant interest \citep{2009BRASP..73..564P,2011Sci...332...69A,2011ApJ...728..122Y,2019SciA....5.3793A,2018PhRvL.120b1101A,2020PhRvL.124u1102A,2021PhRvL.126d1104A,2021PhR...894....1A}. Various models, primarily categorized as sources \citep{2012MNRAS.421.1209T}, propagation \citep{2012ApJ...752L..13T,2015PhRvD..91h3012G,2016ChPhC..40a5101J}, and re-acceleration \citep{2010ApJ...725..184B,2014A&A...567A..33T}, have been proposed to provide explanations. Simultaneously addressing the measured positron excess \citep{2019PhRvL.122d1102A} and the diffuse gamma-ray hardening in the Galactic disk \citep{2012ApJ...750....3A}, the propagation effects with a nearby source are strongly favored \citep{2016ApJ...819...54G,2018PhRvD..97f3008G}. In the case of a nearby source, it is likely that all the indices of primary components are identical, particularly when the energy is higher than several hundred GeV, as the interactions during transportation mainly affect the spectra of lower energy.

The precise measurements and large statistics provided by AMS-02 have unveiled new properties in the nuclei spectra. The primary and secondary components for heavy nuclei from carbon to iron fluxes are estimated by performing fit to the weighted sum of the flux of primary CR oxygen (silicon) and the flux of secondary CR flux boron (fluorine) \citep{2020PhRvL.124u1102A,2021PhRvL.126d1104A,2023PhRvL.130u1002A}. It calibrate their abundance of primary and secondary components independently from models, revealing that primary He-C-O-Fe are distantly different from Ne-Mg-Si-S and there are at least two classes of secondary components \citep{2018PhRvL.121e1103A,2020PhRvL.124u1102A,2021PhRvL.127b1101A,2021PhRvL.126h1102A,2023PhRvL.130u1002A}. The study by \citep{2023RAA....23k5002P} investigated the consistency of injected spectra among different groups of nuclei, assuming spatially uniform propagation, indicating intrinsic differences in the injection spectra.

In this study, we have utilized the spatially-dependent propagation model \citep{2016ApJ...819...54G,2018PhRvD..97f3008G}, which was extended and developed based on the two-halo propagation model proposed by \citep{2012ApJ...752L..13T}, to examine the relative contributions of primary and secondary components in each nuclei spectrum.
The structure of this paper is organized as follows: we initially introduce the spatially-dependent propagation model, followed by the presentation of the results pertaining to each nucleon spectrum and the corresponding abundance outcomes, and ultimately, we provide our concluding remarks.

\section{Methodology}

In this section, we describe the propagation setup that will be used throughout the paper, which is based on the model settings presented in \citep{2016ApJ...819...54G,2018PhRvD..97f3008G}. CR dynamics in the Galaxy is generally described by a differential equation \citep{1964ocr..book.....G,1971ApJ...170..265S,berezinskii1990dogiel,2008JCAP...10..018E} that includes acceleration, loss, and transport terms, described as
\begin{equation}
\begin{aligned}
\frac{\partial \psi}{\partial t}= & Q(\mathbf{r}, p)+\nabla \cdot\left(D_{x x} \nabla \psi-\mathbf{V}_c \psi\right)+\frac{\partial}{\partial p} p^2 D_{p p} \frac{\partial}{\partial p} \frac{1}{p^2} \psi \\
& -\frac{\partial}{\partial p}\left[\dot{p} \psi-\frac{p}{3}\left(\nabla \cdot \mathbf{V}_c \psi\right)\right]-\frac{\psi}{\tau_f}-\frac{\psi}{\tau_r}
\end{aligned}
\end{equation}
where $Q(\mathbf{r}, p)$ is the source function, $D_{xx}$ and $D_{pp}$ are the spatial
diffusion coefficient and diffusion coefficient in the momentum-space, respectively. $\mathbf{V}_c$ is the convection velocity, $\tau_f$ and $\tau_r$ are the characteristic time scales used to describe the fragmentation and radioactive decay.

Spatially dependent diffusion is considered with source-calibrating diffuse coefficient, which is further support by the observation of slow diffusion region around the source \citep{2017Sci...358..911A,2021PhRvL.126x1103A}.
Both CR sources and interstellar medium chiefly spread within the Galactic disk, causing a much slower propagation process close to the Galactic disk ($|z| \leq \xi z_0$ ). While regions far way from the disk ($|z| > \xi z_0$) particles transport as the traditional assumption. The spatial distribution of CR sources \citep{1998ApJ...504..761C} is parameterized as
\begin{equation}
f(r, z)=\left(\frac{r}{r_{\odot}}\right)^{1.25} \exp \left[-\frac{3.87\left(r-r_{\odot}\right)}{r_{\odot}}\right] \exp \left(-\frac{|z|}{z_s}\right),
\end{equation}
where $r_{\odot}$ = 8.5~kpc and $z_s$ = 0.2~kpc.

The propagation coefficient is anti-correlated with the source density and is described as
\begin{equation}
    D_{xx}(r, z, \mathcal R) = D_0 F(r, z) \beta^\eta \left(\frac{\mathcal R}{\mathcal R_0} \right)^{\delta_0 F(r, z)} ~,
\end{equation}
where the function $F(r,z)$ is defined as:
\begin{equation}
   F(r,z) =
   \begin{cases} g(r,z) +\left[1-g(r,z) \right] \left(\dfrac{z}{\xi z_0} \right)^{n} , &  |z| \leq \xi z_0 \\
1 ~, & |z| > \xi z_0
    \end{cases},
\end{equation}
with $g(r,z) = N_m/[1+f(r,z)]$. The distributions of F(r,z) with respect to the radial distance r and vertical height z could be referenced from the work in \citep{2018PhRvD..97f3008G}.

The injection spectrum of sources is assumed to be a broken power-law form, whose power indexes and flux normalization factors are listed in table \ref{tab:inje}.

CR species spectra is obtained by extending DRAGON \citep{2008JCAP...10..018E} to solve the general diffusion-loss transport equation. The corresponding transport parameters are given in table \ref{tab:para}. The force-field approximation \citep{1968ApJ...154.1011G} is adopted to for solar modulation effect.

 \begin{table*}[!htb]
 \setlength{\tabcolsep}{0.8mm}
 \centering
 \caption{Spectral injection parameters and solar modulation energy for each element}\label{tab:inje}%
 \begin{tabular}{ccccc}
 \hline
Element  &  Normalization           & $ \nu_2^{\dagger}$ &  Abundance & $\rm E_{modu}$ \\
         &   $ \rm [GeV^{-1}~m^{-2}~s^{-1}sr^{-1}]$    &          &      &  [GeV/nucleon] \\
 \hline
 H&    $4.35\times10^{-2}$      &   2.45   &   9.41$\times 10^{5}$   &  0.85 \\
 He&       $2.54\times10^{-3}$     &   2.36  &   55000  & 0.75 \\
 C   &      $9.48\times10^{-5}$    &    2.38 &    2050  & 0.7\\
 N   &        $5.09\times10^{-6}$   &    2.38 &     110 & 0.7\\
 O   &     $1.29\times10^{-4}$     &    2.40 &    2800  & 0.6\\
 Ne   &       $1.74\times10^{-5}$     &    2.41 &    377  &0.55\\
 Na   &     $2.31\times10^{-7}$     &    2.41 &     5 &0.75\\
 Mg   &     $2.31\times10^{-5}$     &    2.41 &    500  &0.6\\
 Al   &       $1.99\times10^{-6}$  &    2.41 &     43 &0.45\\
 Si   &     $2.43\times10^{-5}$    &    2.42 &     525 &0.65\\
 S   &      $3.93\times10^{-6}$   &    2.42 &     85 &0.6\\
 Fe   &      $2.59\times10^{-5}$   &    2.43 &     560 &0.8\\
  \hline
    $^\dagger \rm\nu_1 = 2.3, \mathcal{R}_{br} = 6~GeV$.
 \end{tabular}
 \end{table*}

\begin{table*}[!htb]
\caption{Parameters for the SDP propagation model}\label{tab:para}%
\begin{tabular*}{\textwidth}{@{\extracolsep\fill}ccccccc}
\hline
 $ D_0^{\dagger}  $  & $ \delta_{0}$ & $ N_m$ &  $\xi$ & $n$      &  $ v_{A}$       & $ z_0$ \\
 $\rm  [cm^{-2}~s^{-1}]$ &               &        &        &        &  $ [\rm km~s^{-1}]$ & [kpc] \\
 \hline
 $5 \times 10^{28}$& 0.58          &  0.24  & 0.082    & 4.0    & 6               & 5 \\
 \hline
 $^\dagger$Reference rigidity is 4 GV.
\end{tabular*}
\end{table*}

\section{Results}

\begin{figure*}[!htb]
\centering
\includegraphics[width=0.45\textwidth]{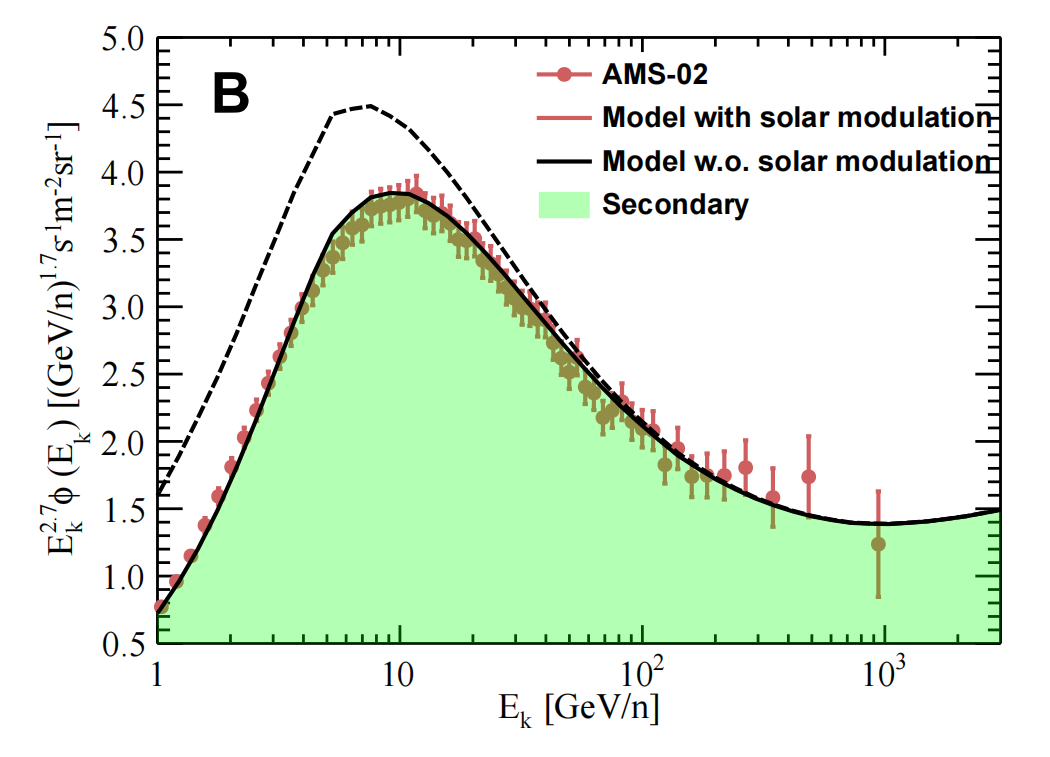}
\includegraphics[width=0.45\textwidth]{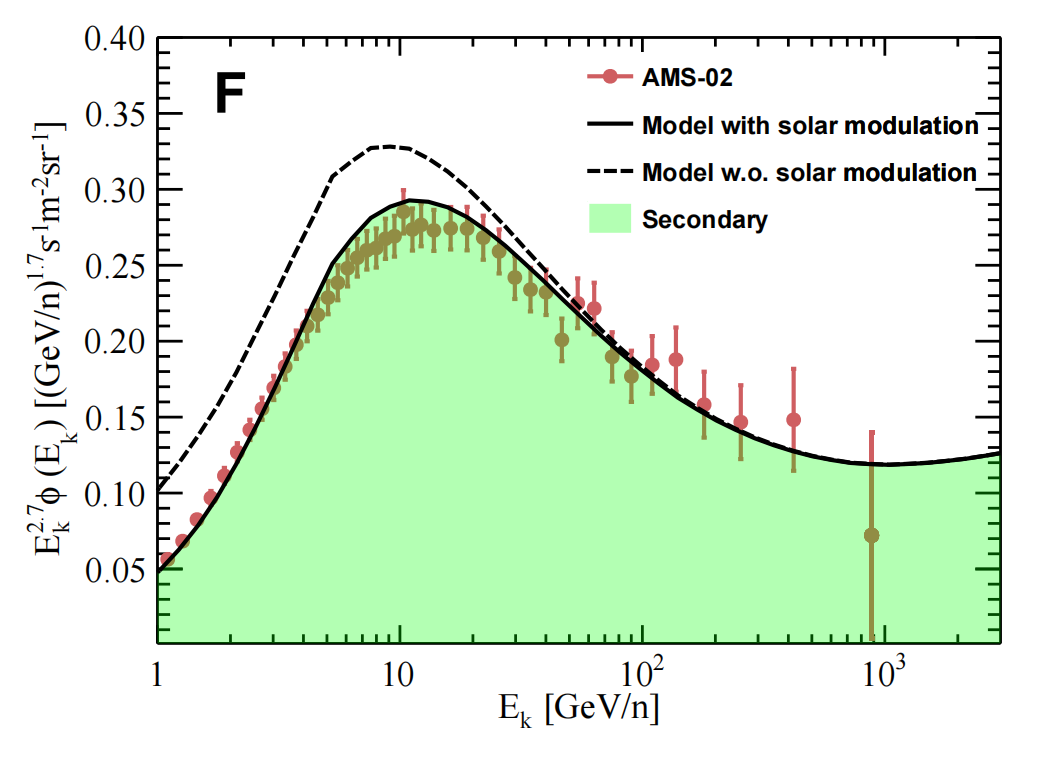}
\caption{
Fluorine and boron CRs, with AMS-02 measurements \citep{2021PhRvL.126h1102A,2018PhRvL.120b1101A}.
}
\label{fig:B}
\end{figure*}

\begin{figure*}[!htb]
\centering
\includegraphics[width=0.45\textwidth]{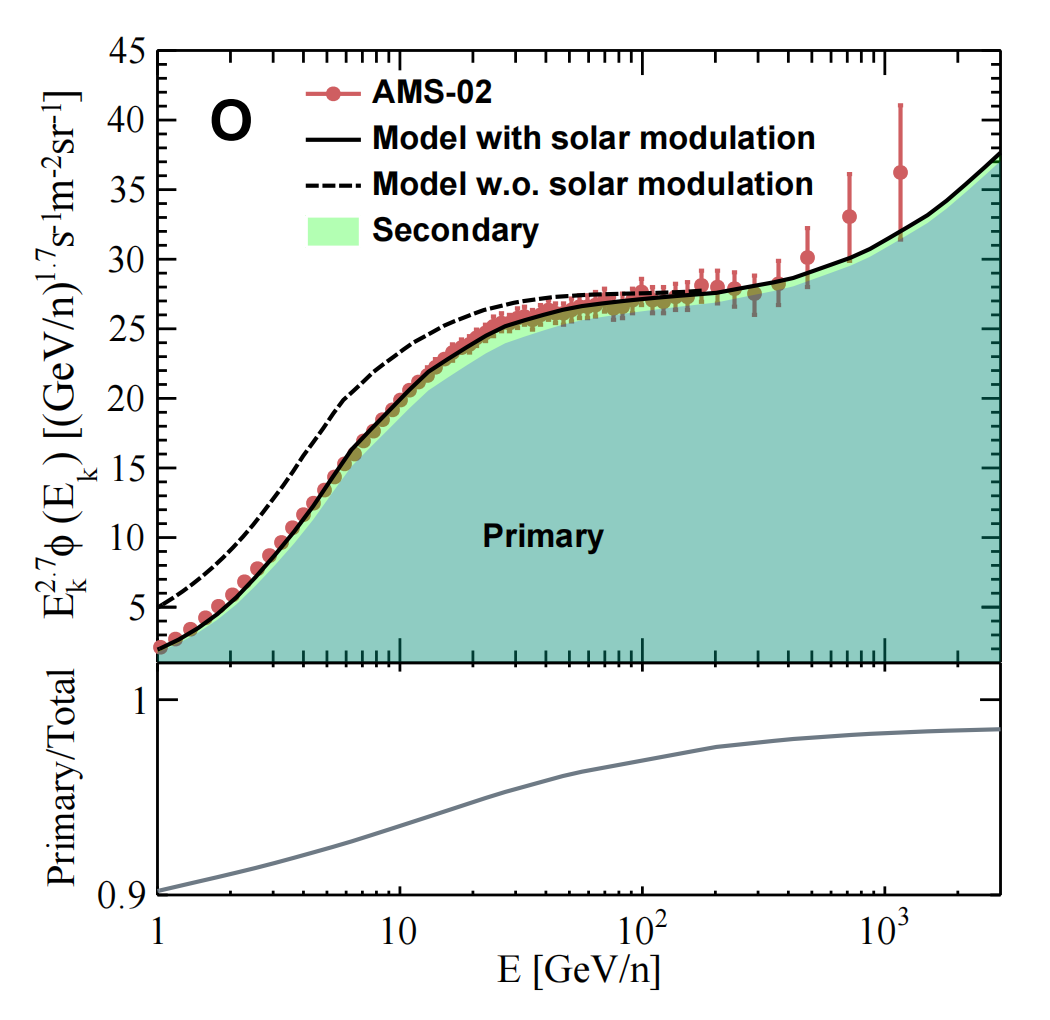}
\includegraphics[width=0.45\textwidth]{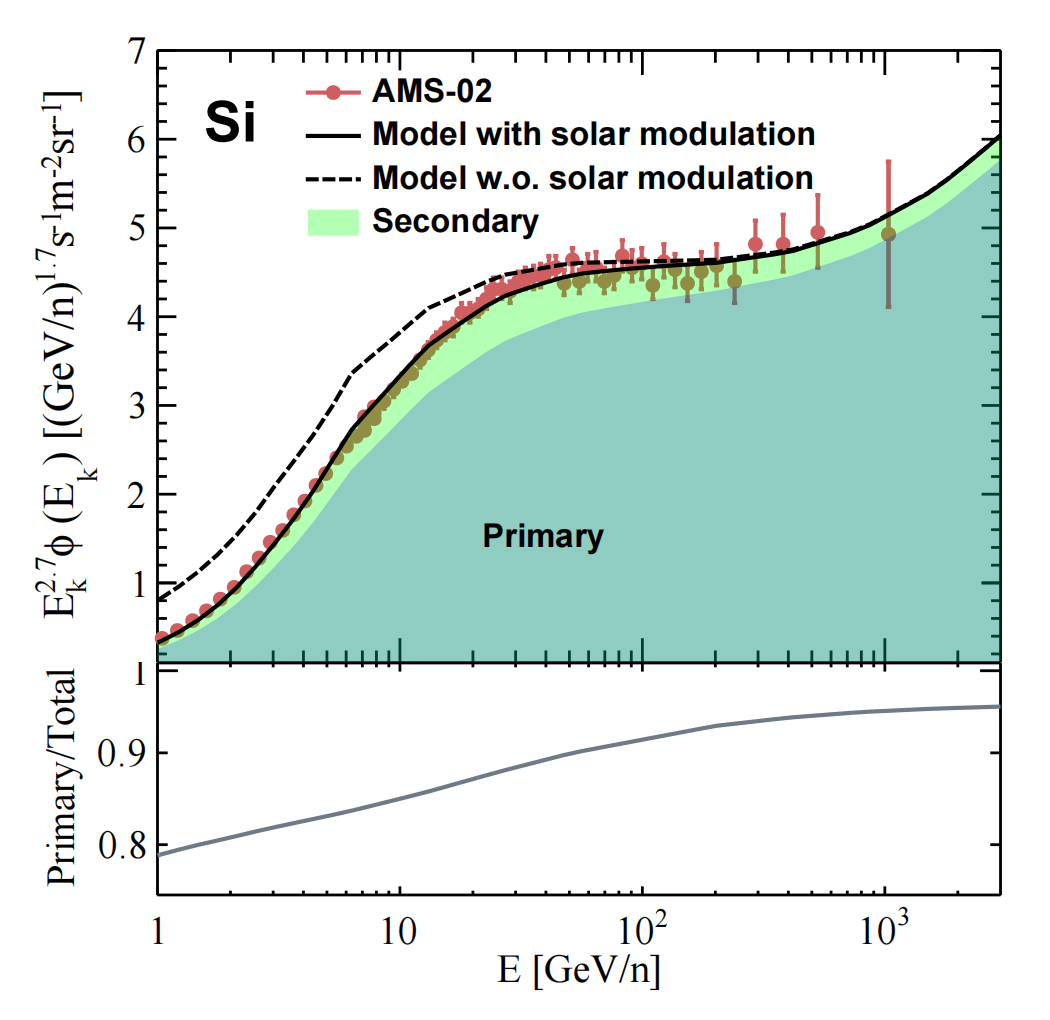}
\caption{Left: The oxygen flux, compared with data from \citep{2017PhRvL.119y1101A}. Right: silicon flux, compare with data from \citep{2020PhRvL.124u1102A}. The primary and secondary component contributions are shown by the dark green and light green shading respectively. The lower panel of each figure presents the primary-to-total flux ratios.
}
\label{fig:si}
\end{figure*}

\begin{figure*}[!htb]
\centering
\includegraphics[width=0.32\textwidth]{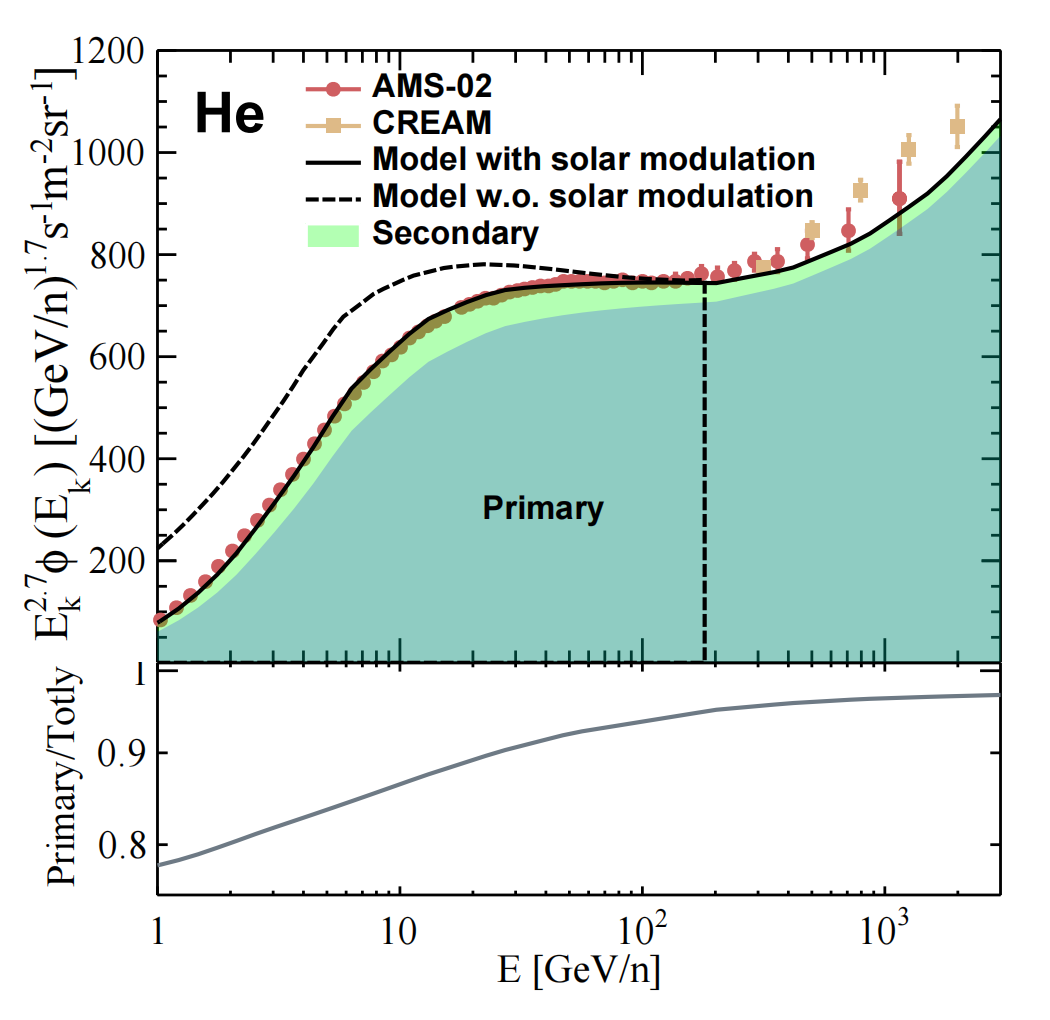}
\includegraphics[width=0.32\textwidth]{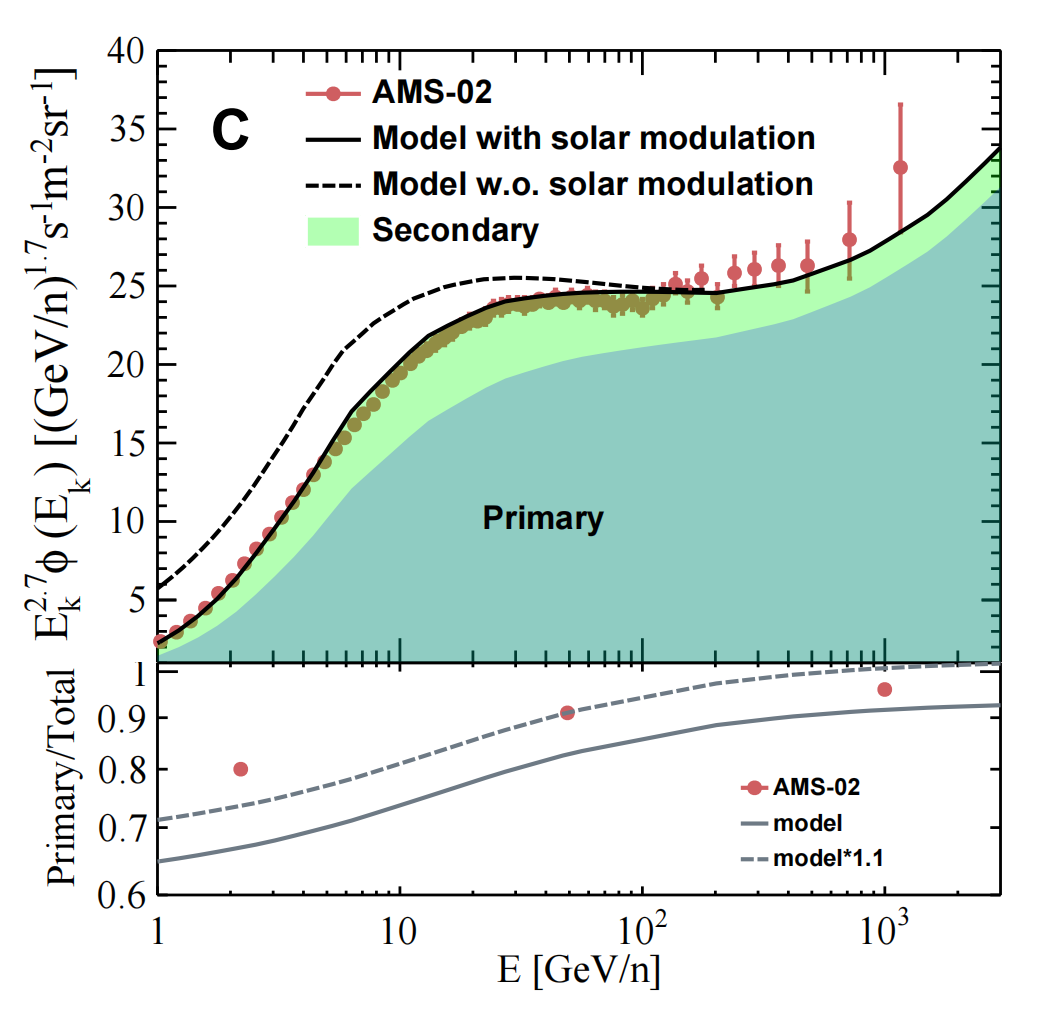}
\includegraphics[width=0.32\textwidth]{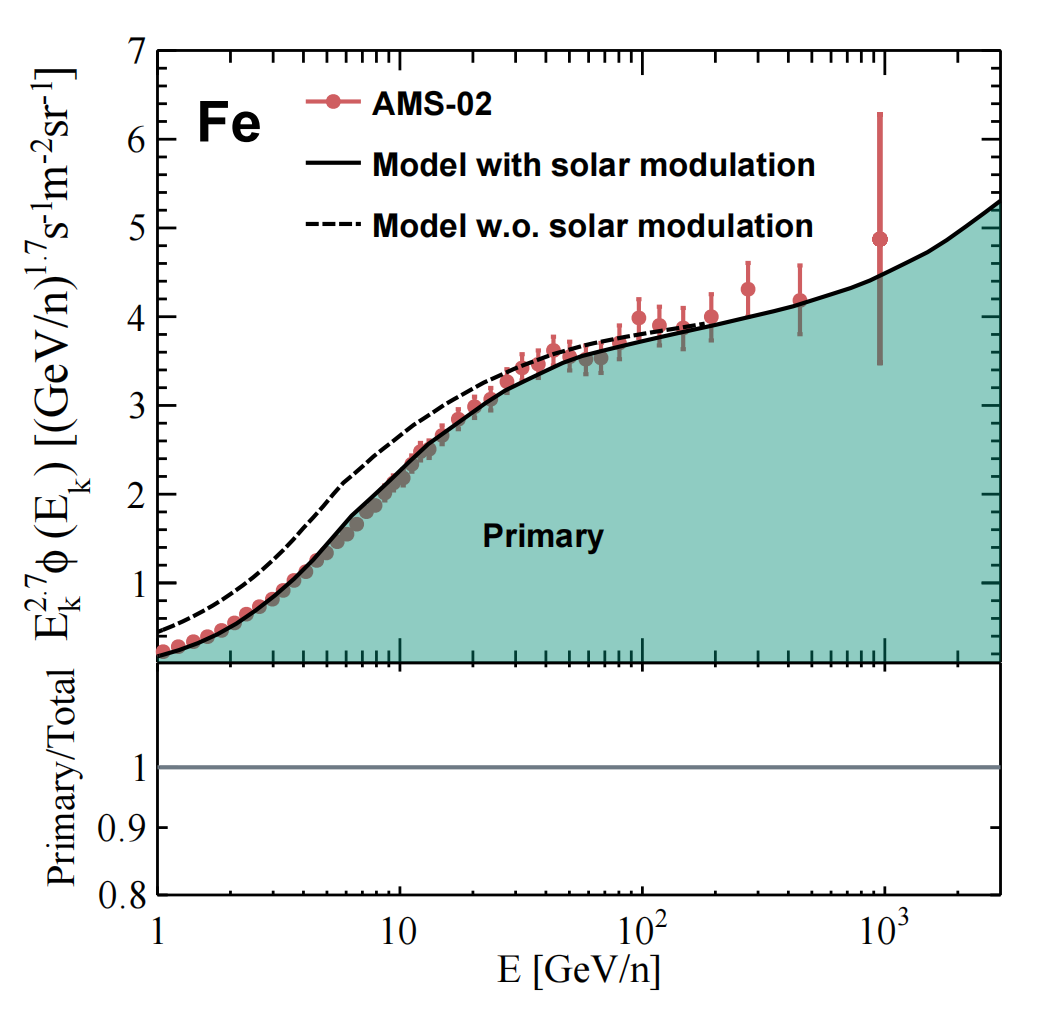}
\includegraphics[width=0.32\textwidth]{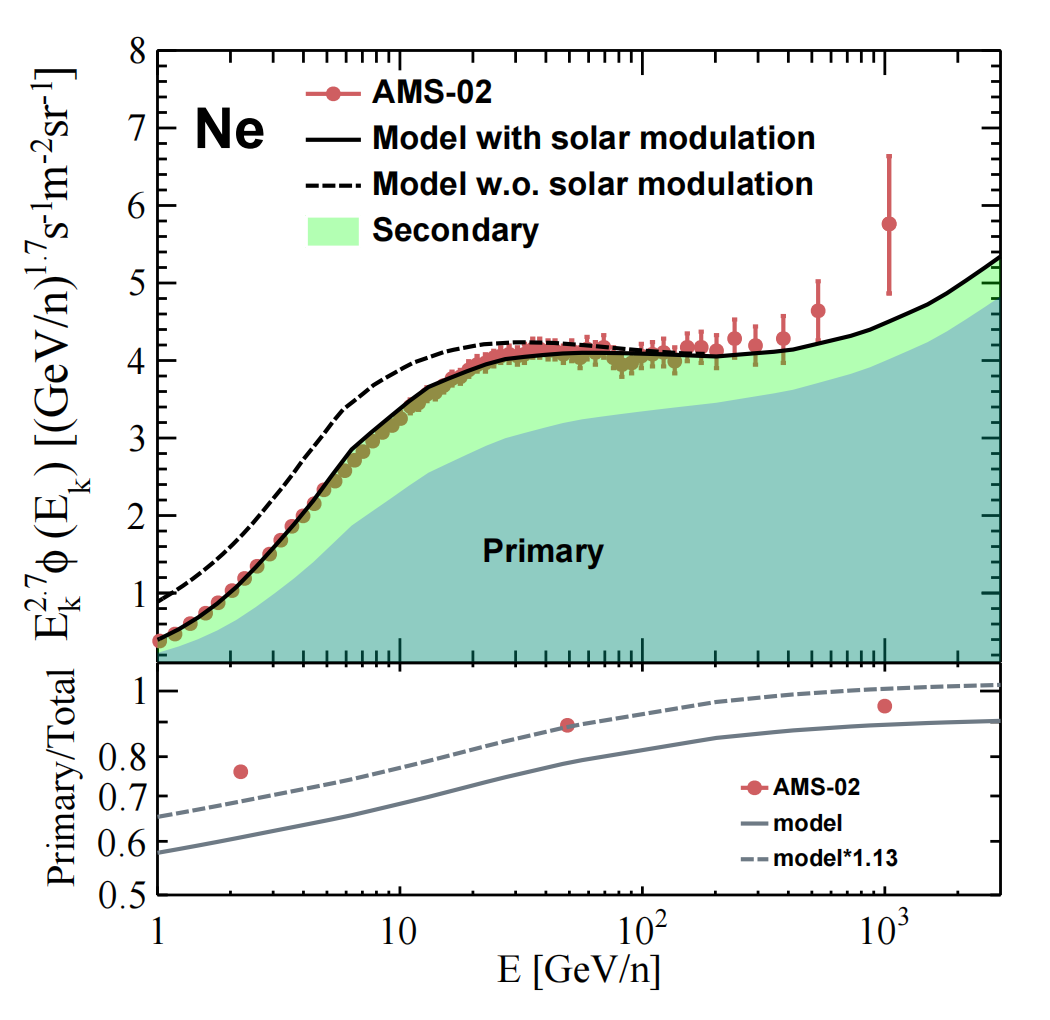}
\includegraphics[width=0.32\textwidth]{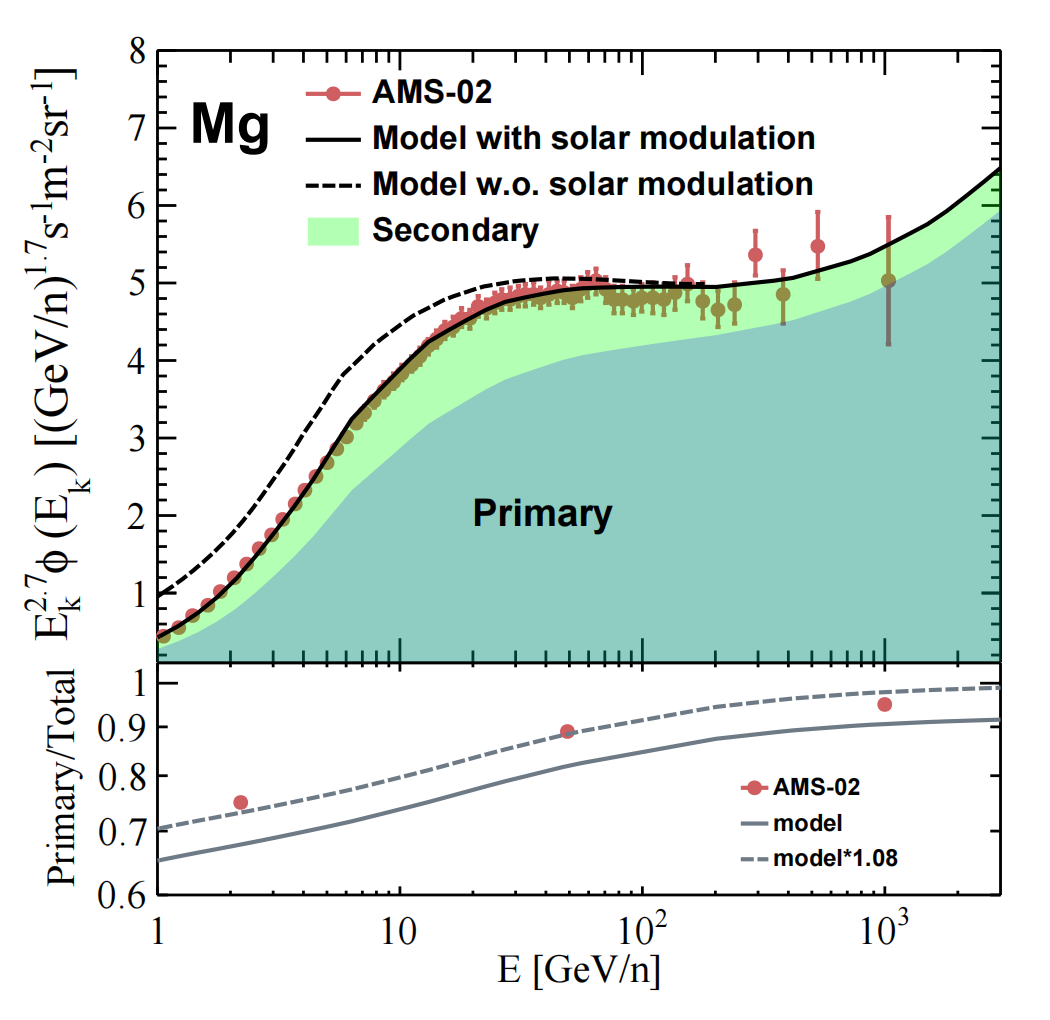}
\includegraphics[width=0.32\textwidth]{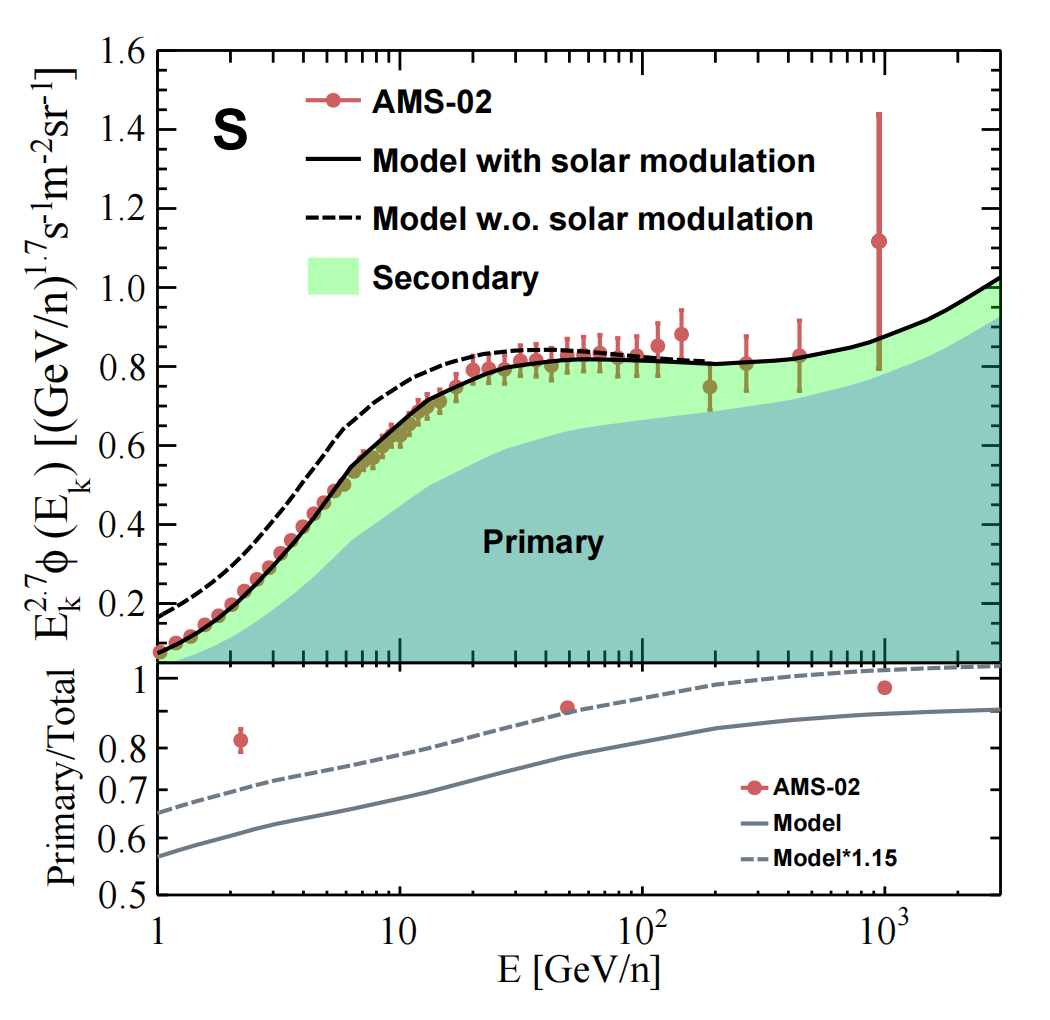}
\caption{
From top to bottom, and from left to right, they are helium, carbon, iron, nickel, magnesium, and sulfur, with AMS-02 measurements \citep{2019PhRvL.123r1102A,2017PhRvL.119y1101A,2021PhRvL.126d1104A,2020PhRvL.124u1102A,2020PhRvL.124u1102A,2023PhRvL.130u1002A}. The primary and secondary component contributions are shown by the dark green and light green shading respectively. The lower panel of each figure presents the primary-to-total flux ratios, compared with ratios from \citep{2023PhRvL.130u1002A}.
}
\label{fig:he}
\end{figure*}



\begin{figure*}[!htb]
\centering
\includegraphics[width=0.32\textwidth]{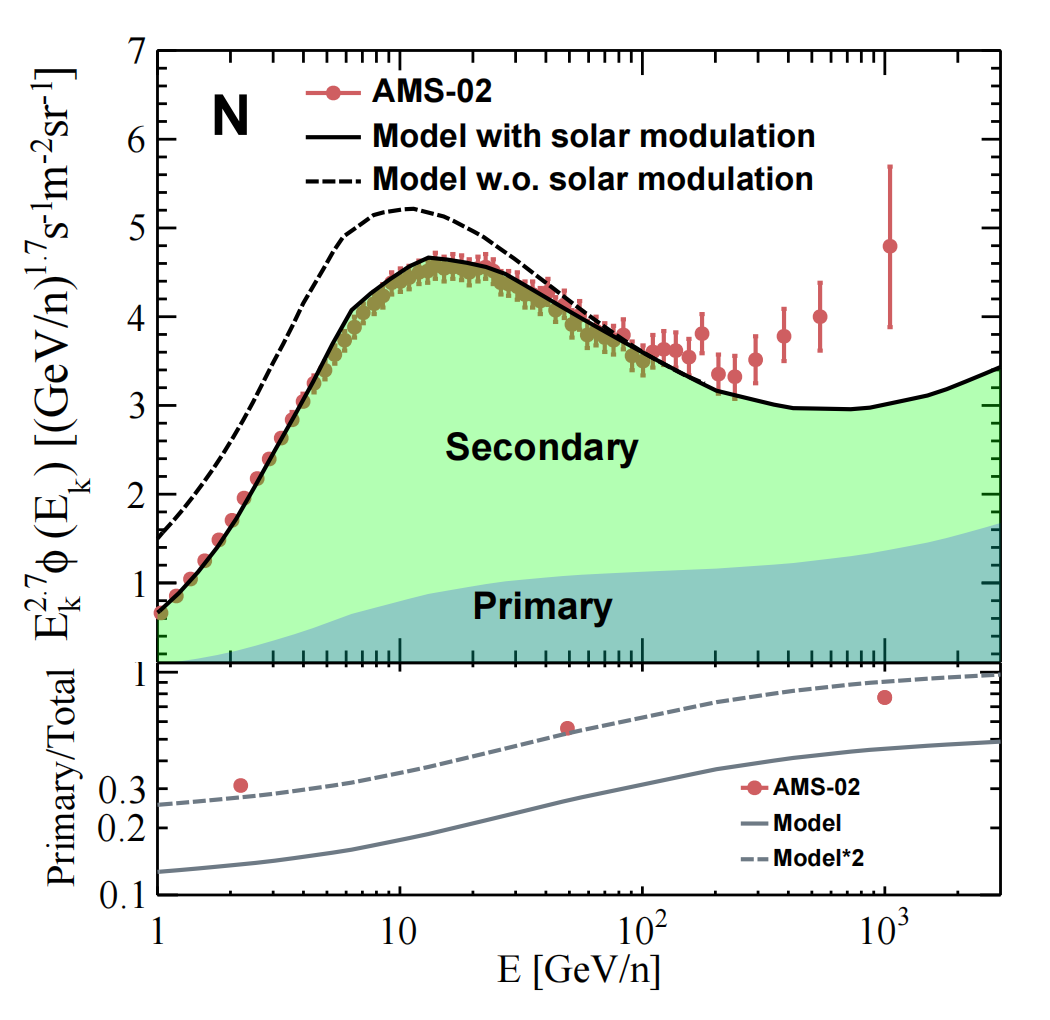}
\includegraphics[width=0.32\textwidth]{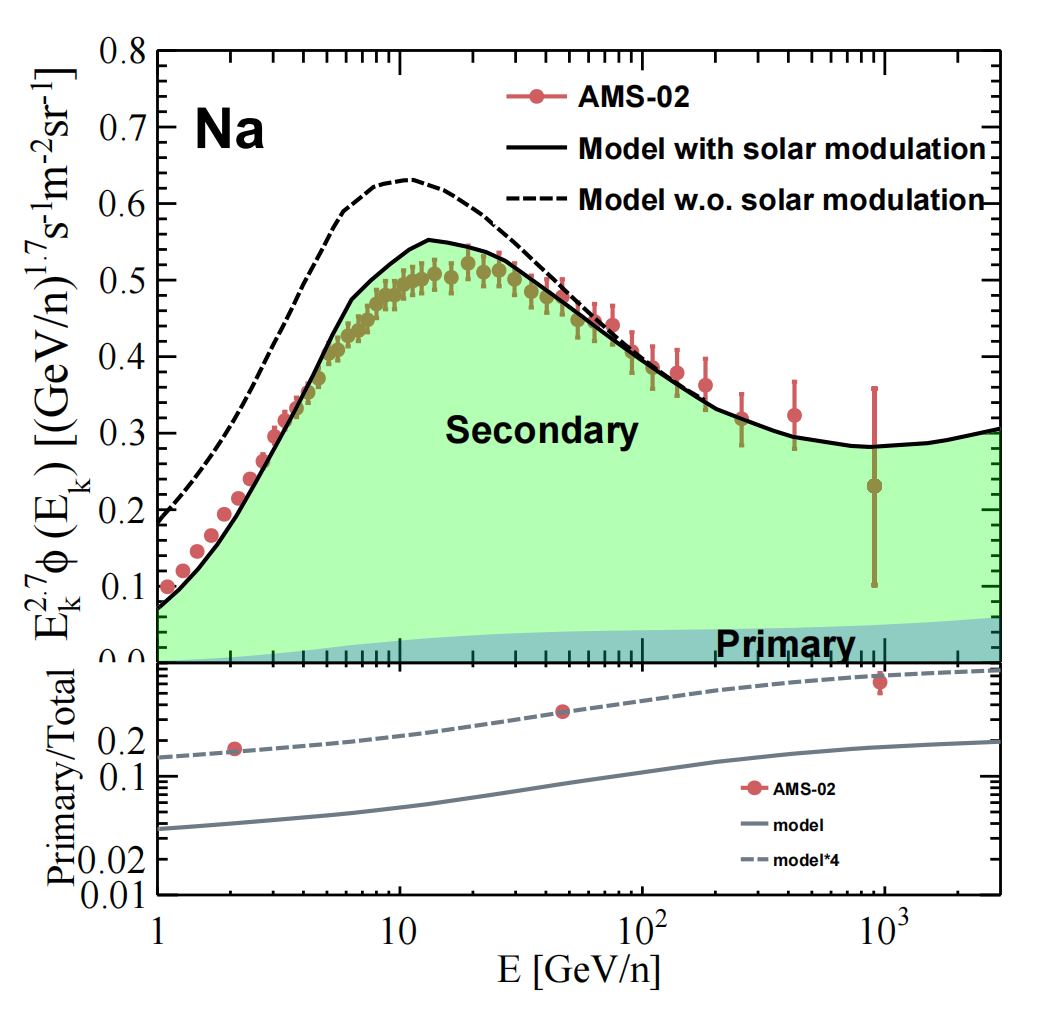}
\includegraphics[width=0.32\textwidth]{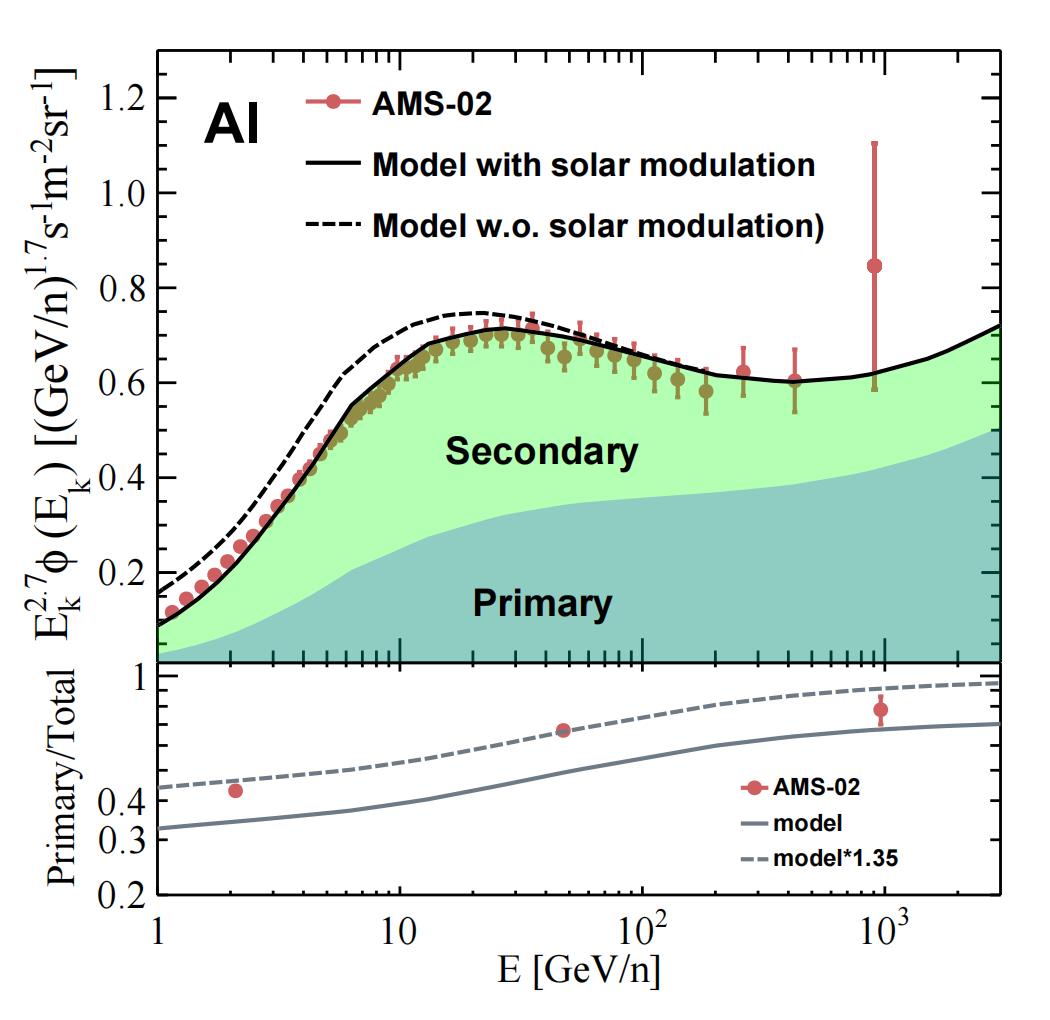}
\caption{
From left to right, they are: nitrogen, sodium, aluminum, with the data from \citep{2021PhRvL.127b1101A}. The lower panel of each figure presents the primary-to-total flux ratios, compared with ratios from \citep{2021PhRvL.127b1101A}.
}
\label{fig:N}
\end{figure*}

\begin{figure*}[!htb]
\centering
\includegraphics[width=0.45\textwidth]{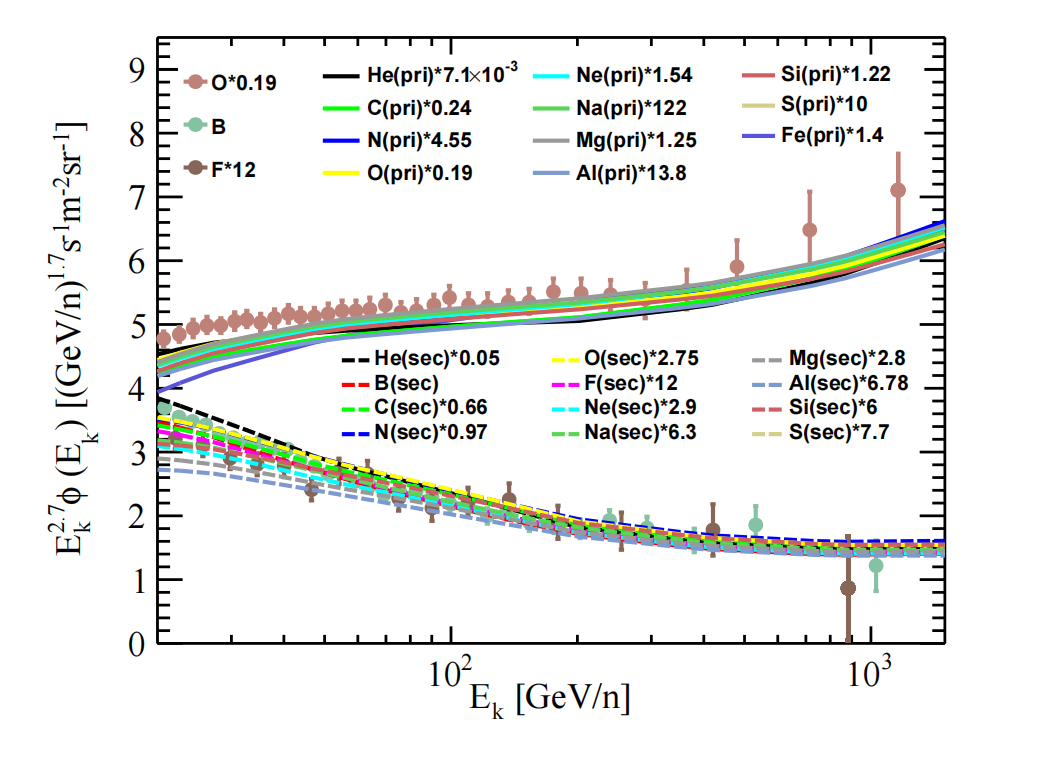}
\includegraphics[width=0.45\textwidth]{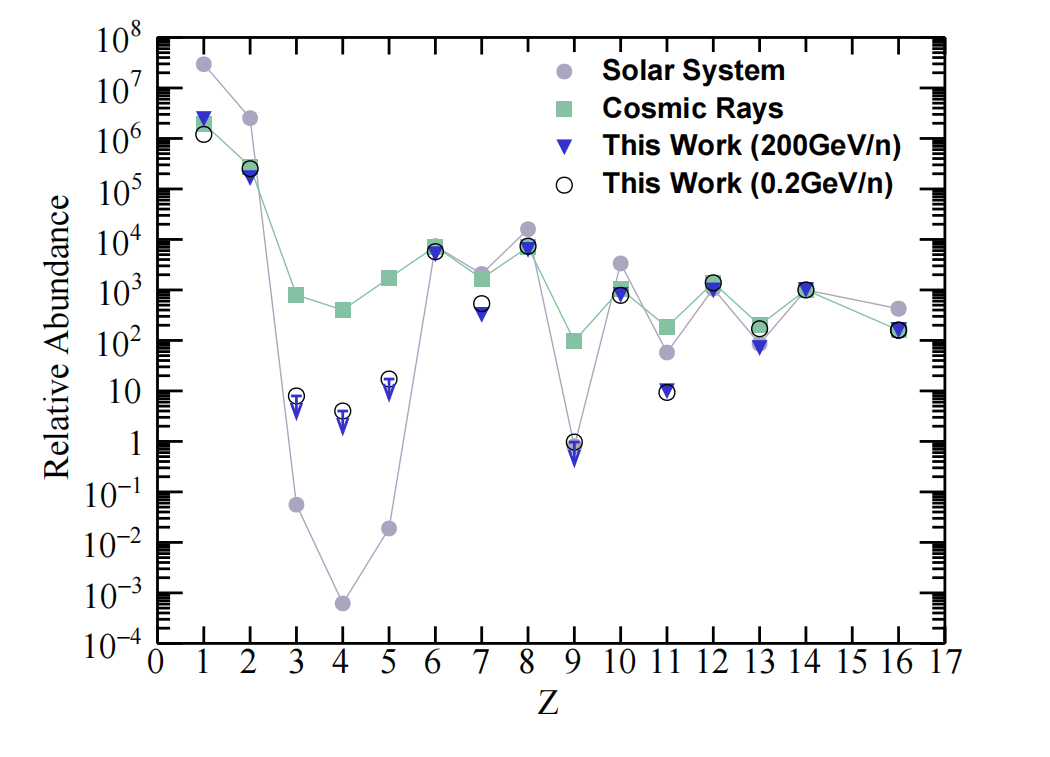}
\caption{ Left: The fluxes of cosmic nuclei from primary and secondary contributions. For display purposes only, the fluxes were rescaled as indicated. Given that fluorine and boron cosmic rays in Figure \ref{fig:B} are purely secondary, and oxygen is almost primary in Figure \ref{fig:si}, the model results are compared with AMS-02 observed fluxes \citep{2021PhRvL.126h1102A,2018PhRvL.120b1101A,2017PhRvL.119y1101A} here as well.  Right: relative abundances of high-energy (200 GeV/nucleon) cosmic rays, compared to the low-energy (0.2 GeV/nucleon) cosmic rays and the present-day solar system, which from \citep{2022PTEP.2022h3C01W}. Abundances are normalised to Si=$10^{3}$.
}
\label{fig:all}
\end{figure*}

The propagated spectra of nuclei (ranging from helium to iron), including their primary and secondary components, are presented as functions of per nucleon kinetic energy. Firstly the boron and fluorine flux are presented in Figure \ref{fig:B}, which are thought purely secondary CRs produced by primary ones during their journal to the Earth. It can be seen that the model-calculated ratios are consistent with the observational data from experiments.

Oxygen and silicon fluxes are given in Figure \ref{fig:si}, as well as their primary, secondary contributions. The primary component represents the injection part and the secondary one mainly stems from the fragmentation of heavier elements. The lower panel of figures illustrate the partitioning of the primary component in relation to the total proton, as a function of energy. It is evident that oxygen is predominantly dominated by the primary component, whereas a relatively small portion (approximately 10\%) at the energy of 10 GeV of the silicon spectra originates from secondary production. The silicon fluxes show a decreasing secondary component and an increasing primary component with increasing energy. In the study of AMS abundance ratios at the source \citep{2023PhRvL.130u1002A}, the oxygen and silicon fluxes are considered to be purely contributed by primary CRs. However, our study reveals that the silicon flux is not as purely primary as that of oxygen.

Figure \ref{fig:he} demonstrates fluxes of primary, secondary, and total flux of helium, carbon, as well as the previously claimed distinct classes nickel, magnesium, and sulfur, revealing that all their primary-to-total flux ratios are similar to each other. Additionally, the solid and dashed lines in the lower panel of each figure represent our model-calculated ratios and our model-calculated ratios times a factor to match the AMS abundance ratios at the source \citep{2023PhRvL.130u1002A}, respectively. It is evident that out model-calculated ratios are all lower than those based on experimental data, and the reason for this is that the observed silicon flux contains not negligible secondaries.

The nitrogen, sodium, and aluminum fluxes are displayed from left to right in Figure \ref{fig:N}. These ratios are notably distinct from those shown in Figure \ref{fig:he} due to their substantial secondary components. Additionally, the primary-to-total flux ratios in the lower panel of each figure indicate significant differences between our calculations and results based on observational secondary and primary components. This suggests that there may be a greater secondary contribution to the spectra of these secondary-dominated nuclei than previously estimated \citep{2021PhRvL.127b1101A}.

The left panel of Figure \ref{fig:all} displays the primary and secondary components of nuclei from helium to iron. The fluxes are rescaled as indicated for display purposes only. It can be observed that all primary components are mostly the same, which is also evident from the injected spectral index listed in table \ref{tab:inje}. However, there are marginal differences between the secondary components in the lower energy range, mainly due to the cross-section differences in secondary production. The right panel of Figure \ref{fig:all} presents the relative abundances of high-energy (200 GeV/nucleon) cosmic rays, normalised to Si=$10^{3}$ and compared to the low-energy (0.2 GeV/nucleon) cosmic rays and the present-day solar system from \citep{2022PTEP.2022h3C01W}. It is evident that there are significant differences for each element compared to those in the solar system, especially for the Z-odd ones.

Here we investigate the spectra of different nuclei primarily based on the AMS-02 data, as the target energy range in this study exceeds tens of GeV, where it is free from solar modulation. The model calculations compared with measurements for each nucleus at low energies outside the solar system by Voyager-1 \citep{2016ApJ...831...18C} are presented in the appendix \ref{secA1}. It is evident that there are significant deviations between the model calculations without solar modulation and Voyager's observations. The origin of this discrepancy may be due to Voyager still being within the influence of the solar magnetic field, or it could be attributed to the accuracy of solar modulation in this work. These two aspects will be addressed in our future work, although they do not impact the conclusions drawn in this paper.


\section{Summary}
This work is aimed at understanding the primary and secondary components of each CR species recently observed by AMS. We took advantage of SDP propagation model, tracing the spectra from originate from sources and production during the transportation. We found that boron and fluorine are purely secondaries while the silicon spectra is not as pure primary as the oxygen. The primary component of CR species (He-C-O-Ne-Mg-Si-S-Fe) are the same class, N-Na-Al are secondary-dominated. All primary component are increasing with energy. When particle energies are above TeV, diffuse propagation dominates and particle interaction is negligible. If they were one group, they would stay together with higher energy. Future more precise measurements above TeV could test if there are significant spectral differences and validate our model calculations. The primary abundance of CR nuclei presented differs from that of the solar system. This clean data set of primary and secondary component could help us to check the consistency between the observed data and the CR model.

\begin{acknowledgments}
This work is supported by the National Natural Science Foundation of China (Nos.12275279, 12373105, 12320101005) and the China Postdoctoral Science Foundation (No.2023M730423).
\end{acknowledgments}


\appendix

\section{Comparison between model data and the local interstellar observations}\label{secA1}
\setcounter{table}{0}
\setcounter{figure}{0}
\renewcommand{\thetable}{A\arabic{table}}
\renewcommand{\thefigure}{A\arabic{figure}}
The model calculations extended to MeV per nucleon, compared with measurements for each nucleus at low energies outside the solar system by Voyager-1 \citep{2016ApJ...831...18C}, are presented in Figures \ref{fig_A:B}, \ref{fig_A:he}, and \ref{fig_A:N}.

\begin{figure*}[!htb]
\centering
\includegraphics[width=0.45\textwidth]{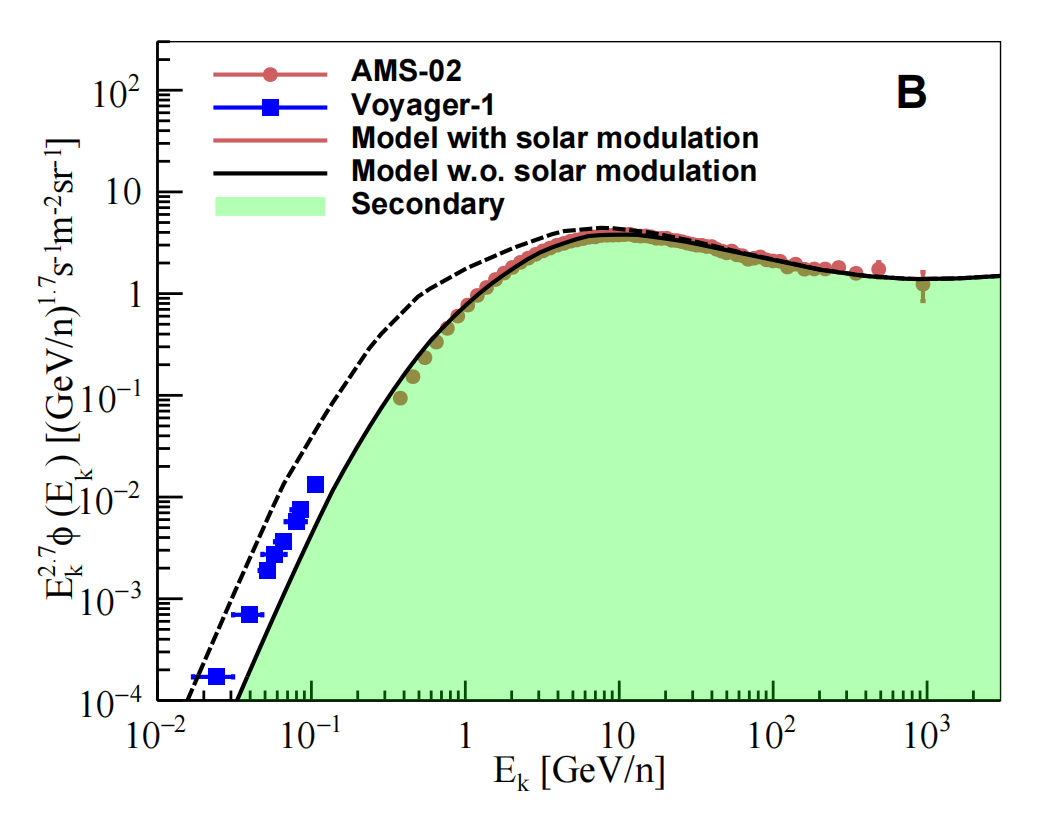}
\includegraphics[width=0.45\textwidth]{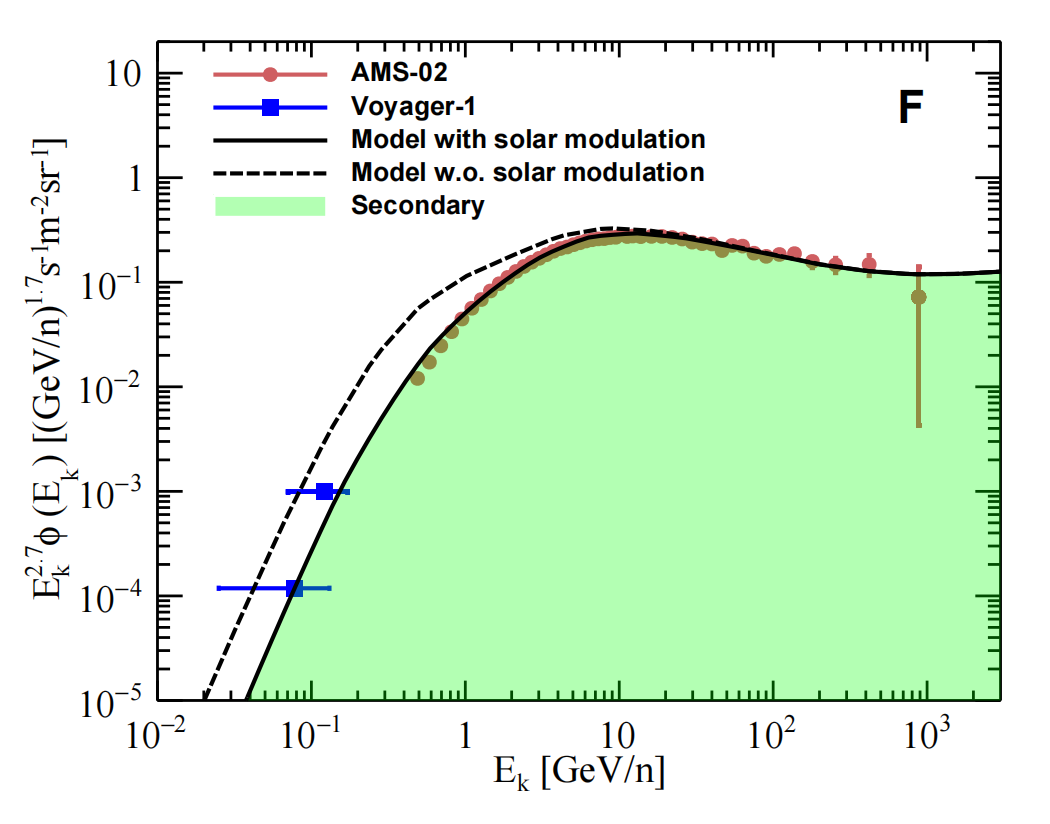}
\includegraphics[width=0.45\textwidth]{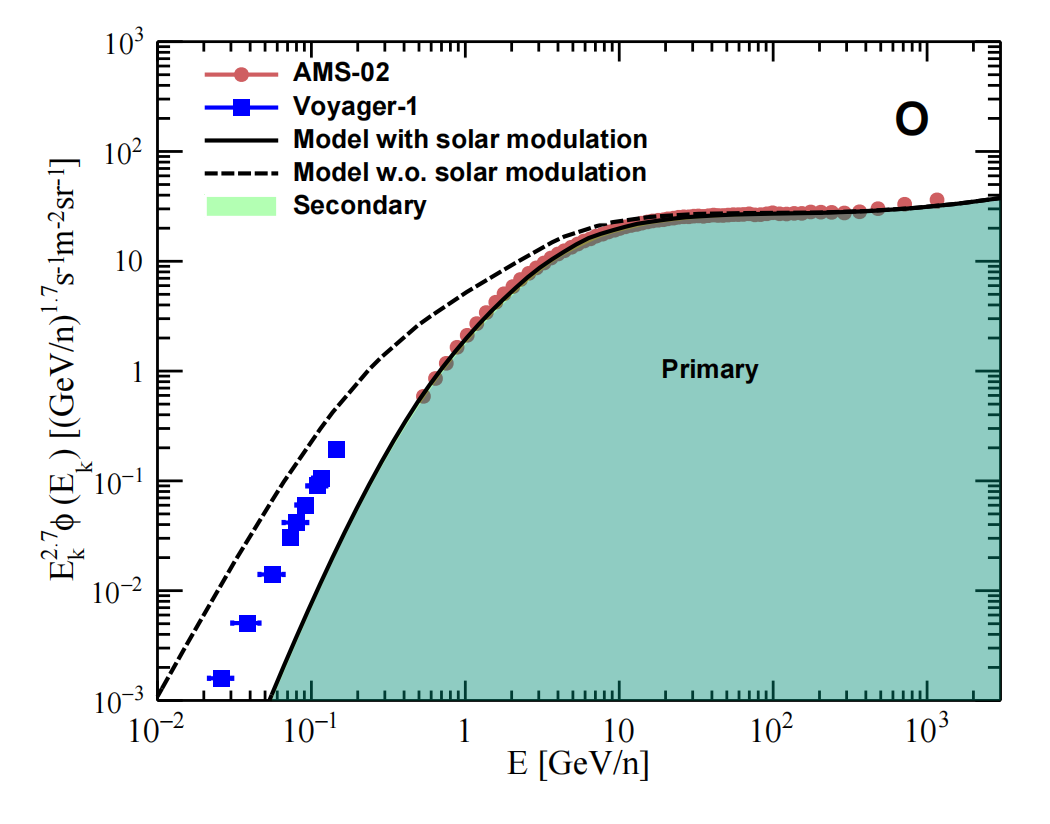}
\includegraphics[width=0.45\textwidth]{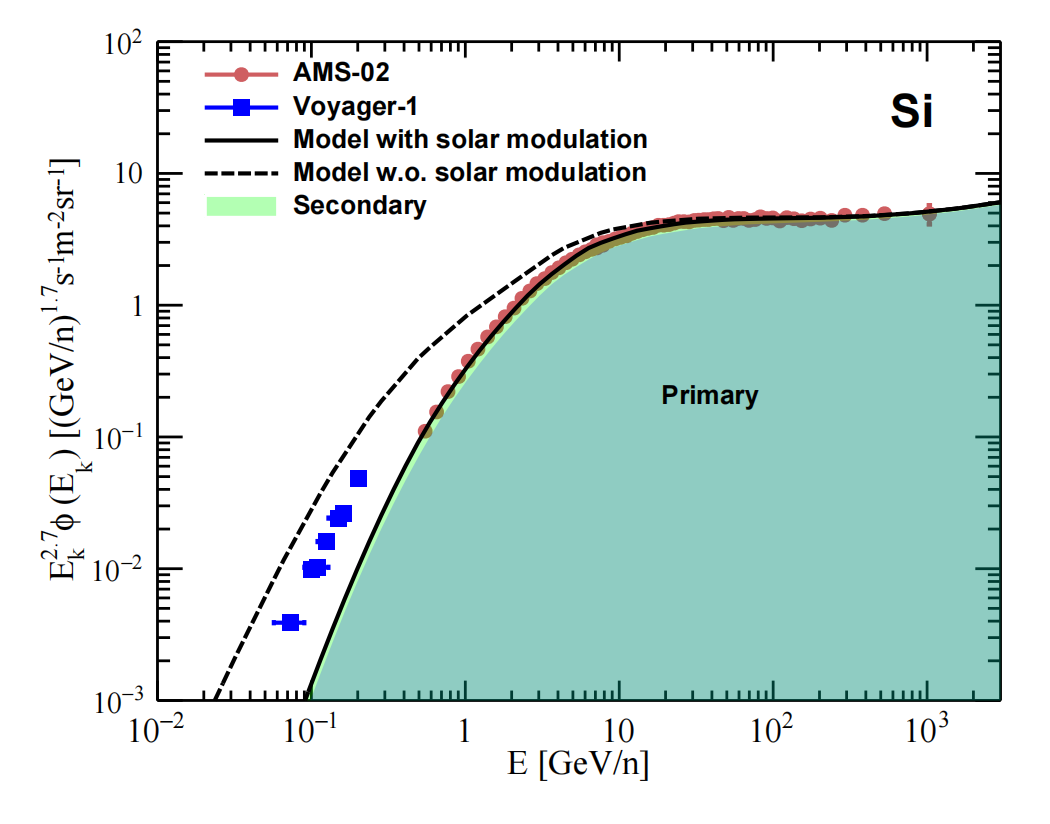}
\caption{
Fluorine, boron, oxygen, and silicon CRs, with AMS-02 \citep{2021PhRvL.126h1102A,2018PhRvL.120b1101A,2017PhRvL.119y1101A,2020PhRvL.124u1102A} and the Voyager-1 \citep{2016ApJ...831...18C} observations. The model data in this figure is the same as in Figure \ref{fig:B} and Figure \ref{fig:si}, except that in this figure, the low-energy range is extended to tens of MeV per nucleon for comparison with Voyager's observations.
}
\label{fig_A:B}
\end{figure*}


\begin{figure*}[!htb]
\centering
\includegraphics[width=0.32\textwidth]{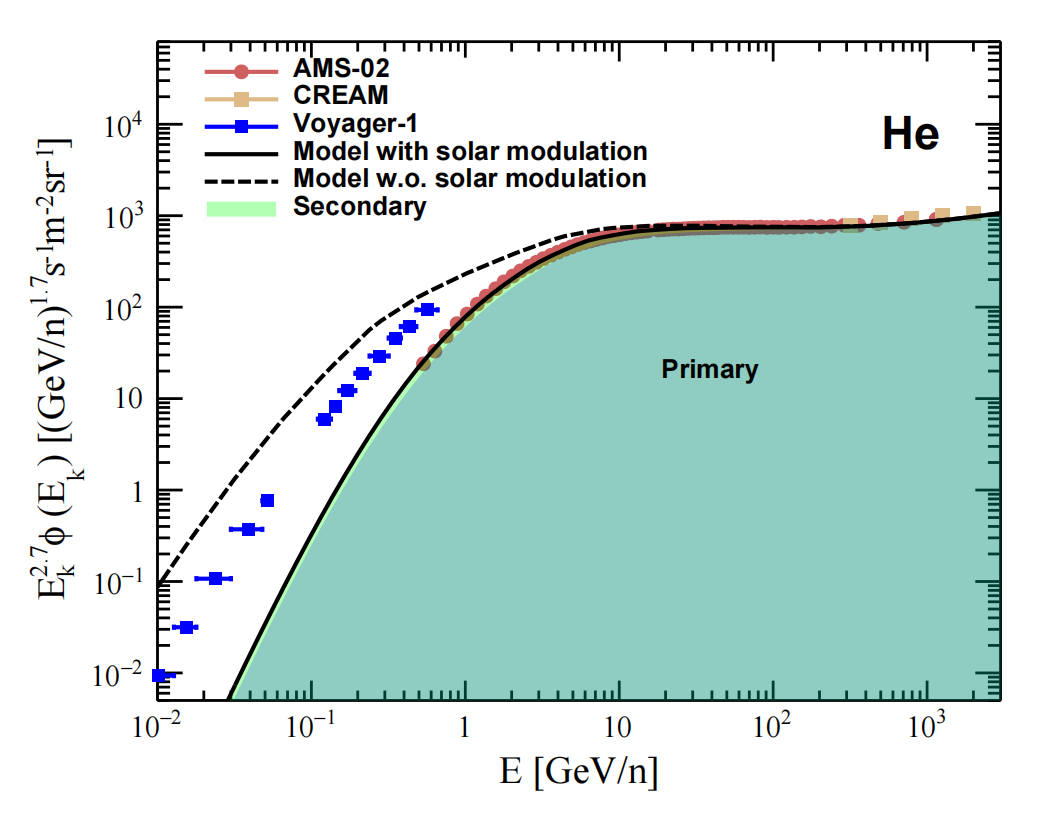}
\includegraphics[width=0.32\textwidth]{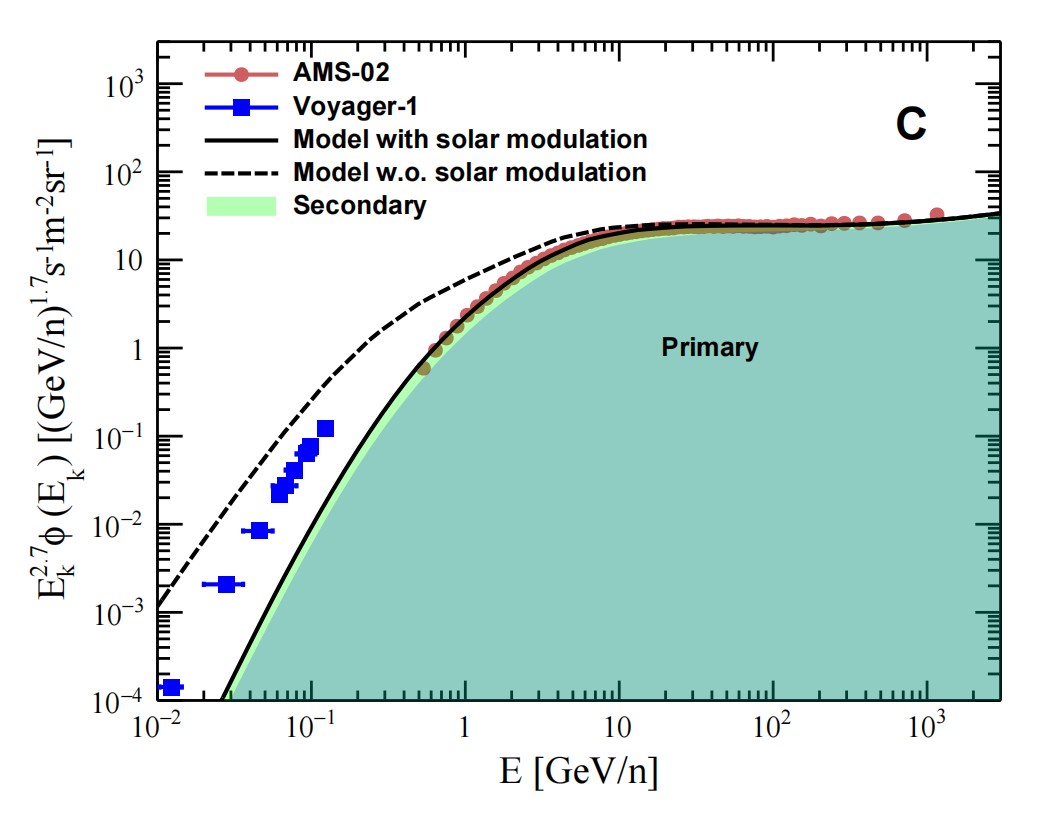}
\includegraphics[width=0.32\textwidth]{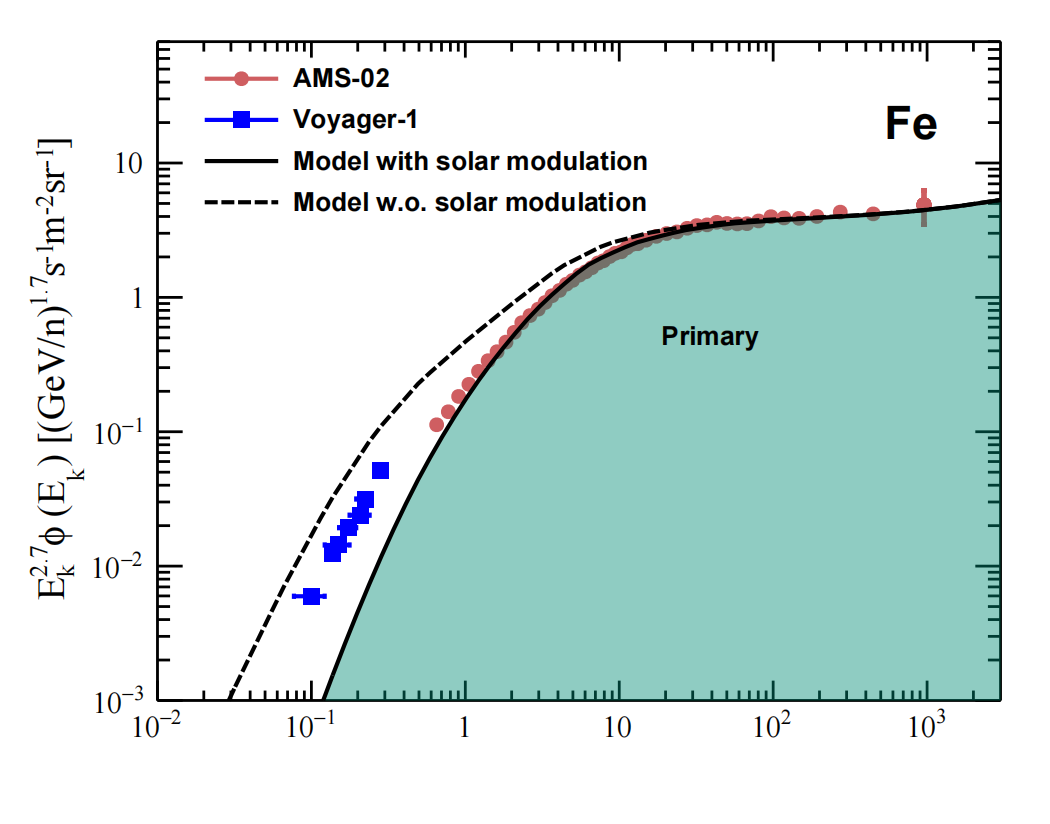}
\includegraphics[width=0.32\textwidth]{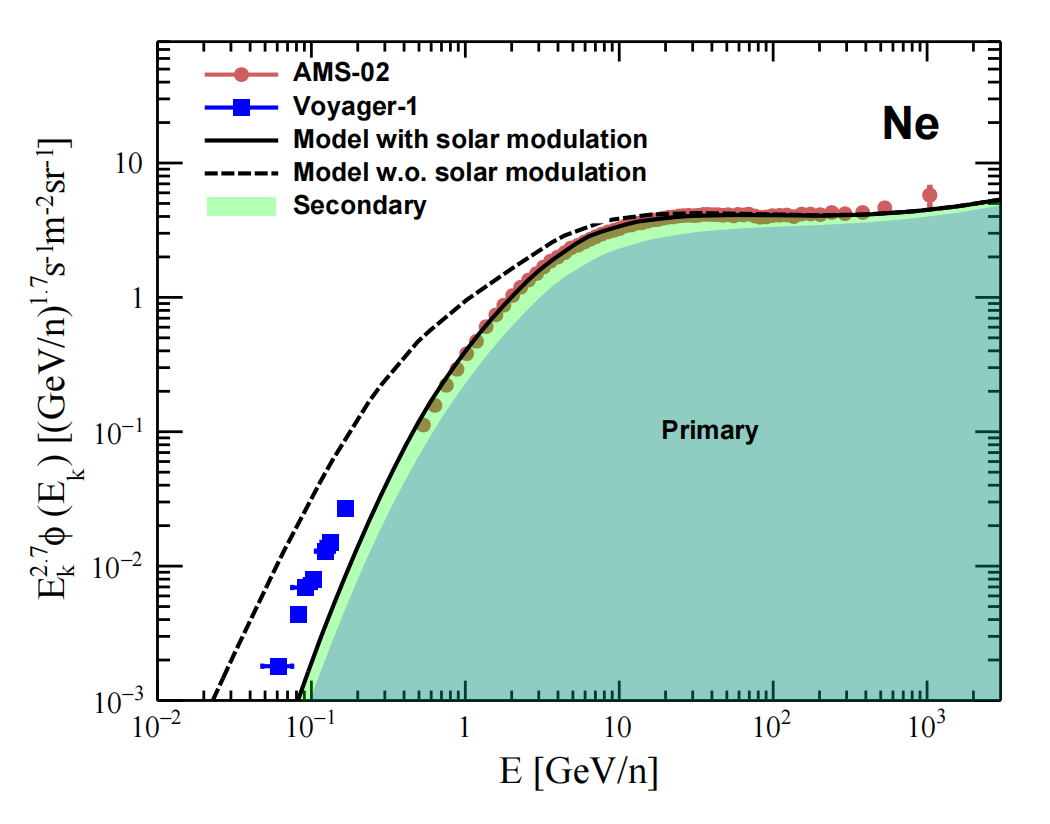}
\includegraphics[width=0.32\textwidth]{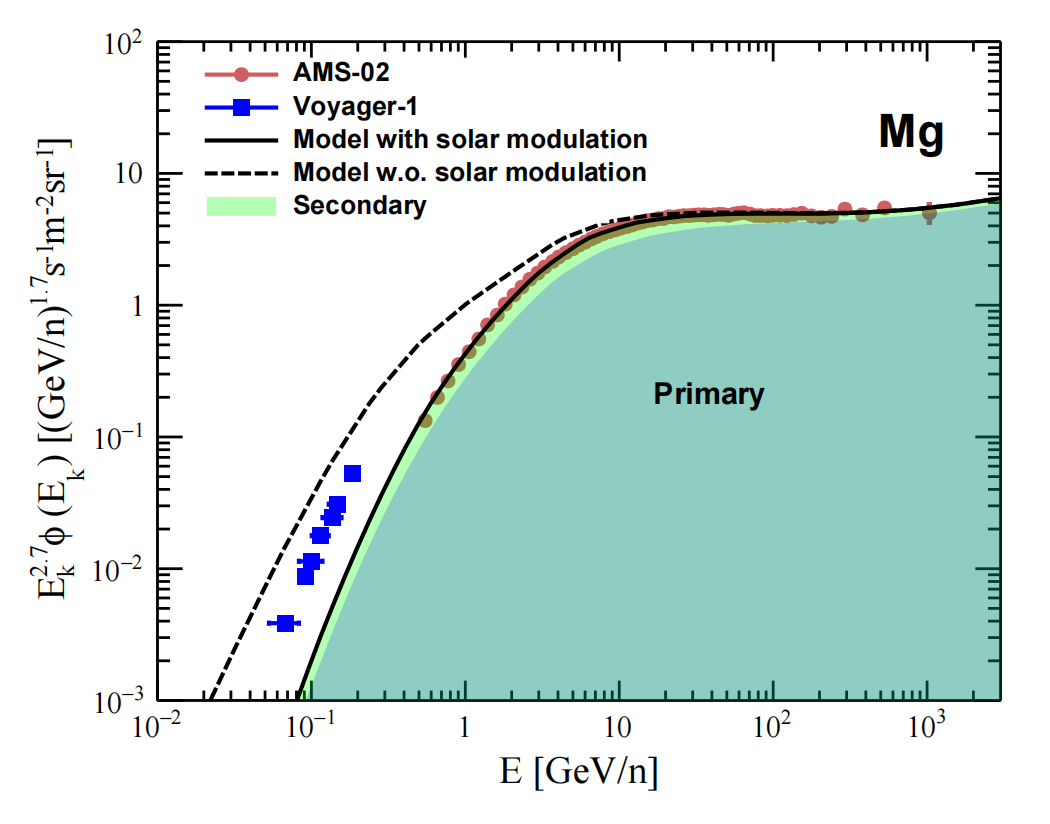}
\includegraphics[width=0.32\textwidth]{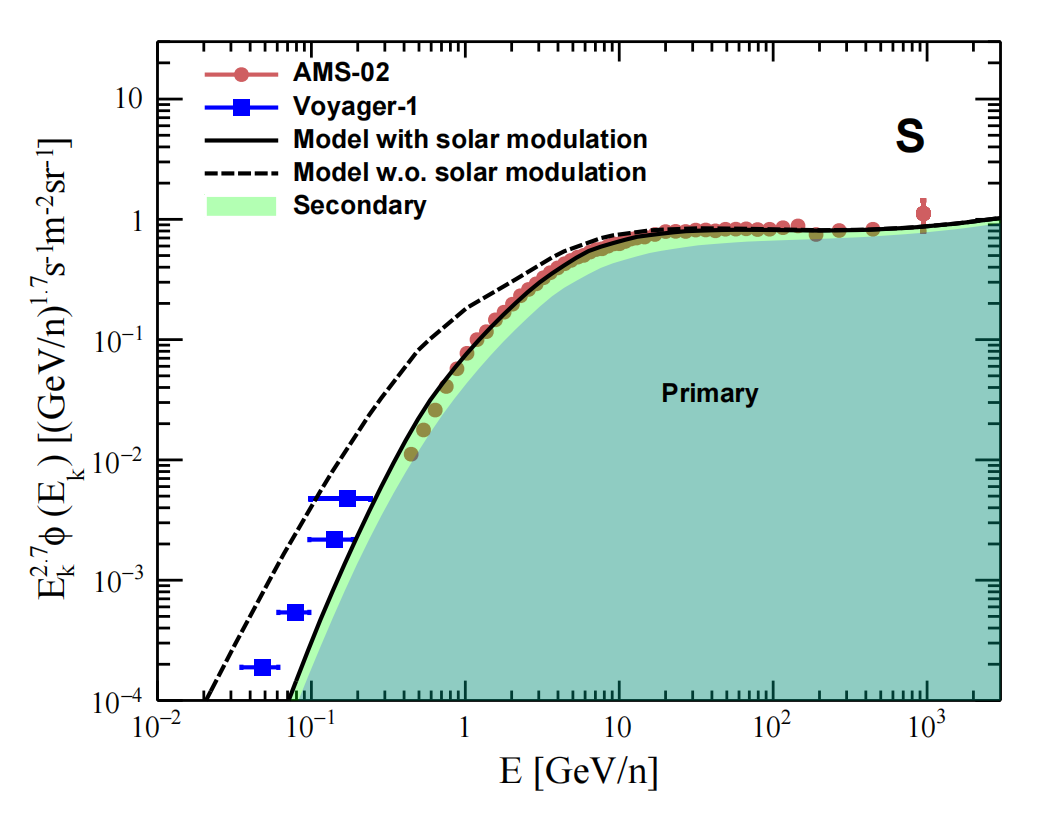}
\caption{
From top to bottom, and from left to right, they are helium, carbon, iron, nickel, magnesium, and sulfur, with AMS-02 \citep{2019PhRvL.123r1102A,2017PhRvL.119y1101A,2021PhRvL.126d1104A,2020PhRvL.124u1102A,2020PhRvL.124u1102A,2023PhRvL.130u1002A} and Voyager-1 \citep{2016ApJ...831...18C} observations. The model data in this figure is the same as in Figure \ref{fig:he}, except that in this figure, the low-energy range is extended to tens of MeV per nucleon for comparison with Voyager's observations.
}
\label{fig_A:he}
\end{figure*}



\begin{figure*}[!htb]
\centering
\includegraphics[width=0.32\textwidth]{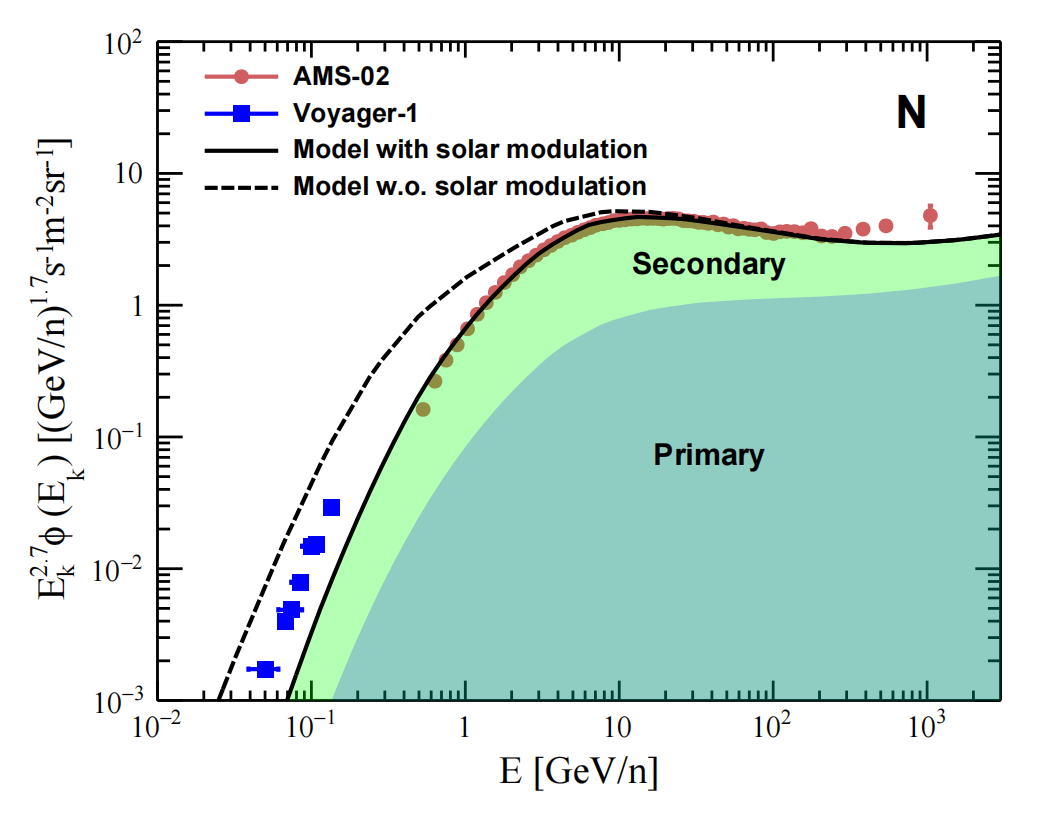}
\includegraphics[width=0.32\textwidth]{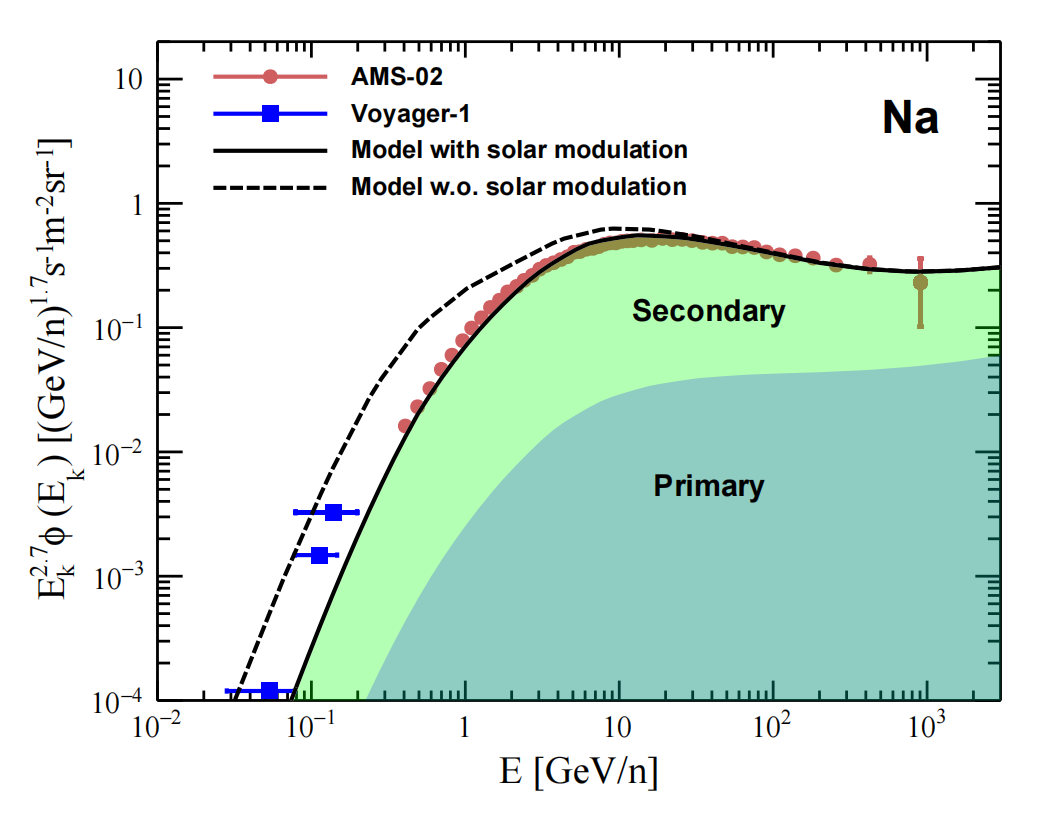}
\includegraphics[width=0.32\textwidth]{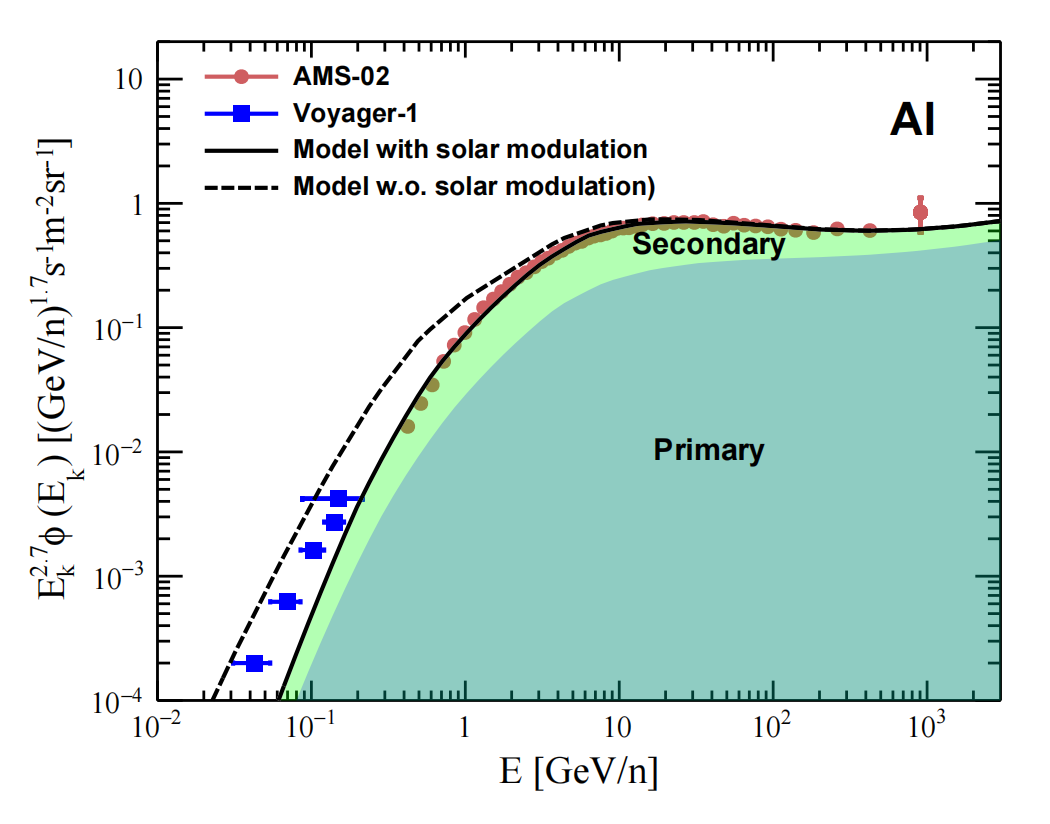}
\caption{
From left to right, they are: nitrogen, sodium, aluminum, with the data from AMS-02 \citep{2021PhRvL.127b1101A} and Voyager-1 \citep{2016ApJ...831...18C} observations. The model data in this figure is the same as in Figure \ref{fig:N}, except that in this figure, the low-energy range is extended to tens of MeV per nucleon for comparison with Voyager's observations.
}
\label{fig_A:N}
\end{figure*}

\bibliographystyle{unsrt_update}
\bibliography{apssamp}

\begin{thebibliography}{10}

\bibitem{1934PNAS...20..254B}
W.~{Baade} and F.~{Zwicky}.
\newblock {On Super-novae}.
\newblock {\em Proceedings of the National Academy of Science}, 20(5):254--259,
  May 1934.

\bibitem{2022JCAP...10..081B}
Olivia~Meredith {Bitter} and Dan {Hooper}.
\newblock {Constraining the local pulsar population with the cosmic-ray
  positron fraction}.
\newblock {\em \jcap}, 2022(10):081, October 2022.

\bibitem{2015ARA&A..53..199G}
Isabelle~A. {Grenier}, John~H. {Black}, and Andrew~W. {Strong}.
\newblock {The Nine Lives of Cosmic Rays in Galaxies}.
\newblock {\em \araa}, 53:199--246, August 2015.

\bibitem{2019ApJS..245...30L}
Richard~E. {Lingenfelter}.
\newblock {The Origin of Cosmic Rays: How Their Composition Defines Their
  Sources and Sites and the Processes of Their Mixing, Injection, and
  Acceleration}.
\newblock {\em \apjs}, 245(2):30, December 2019.

\bibitem{2009BRASP..73..564P}
A.~D. {Panov}, J.~H. {Adams}, H.~S. {Ahn}, et~al.
\newblock {Energy spectra of abundant nuclei of primary cosmic rays from the
  data of ATIC-2 experiment: Final results}.
\newblock {\em Bulletin of the Russian Academy of Sciences, Physics},
  73:564--567, June 2009.

\bibitem{2011Sci...332...69A}
O.~{Adriani}, G.~C. {Barbarino}, G.~A. {Bazilevskaya}, et~al.
\newblock {PAMELA Measurements of Cosmic-Ray Proton and Helium Spectra}.
\newblock {\em Science}, 332:69--, April 2011.

\bibitem{2011ApJ...728..122Y}
Y.~S. {Yoon}, H.~S. {Ahn}, P.~S. {Allison}, et~al.
\newblock {Cosmic-ray Proton and Helium Spectra from the First CREAM Flight}.
\newblock {\em \apj}, 728:122, February 2011.

\bibitem{2019SciA....5.3793A}
Q.~{An}, R.~{Asfandiyarov}, P.~{Azzarello}, et~al.
\newblock {Measurement of the cosmic ray proton spectrum from 40 GeV to 100 TeV
  with the DAMPE satellite}.
\newblock {\em Science Advances}, 5(9):eaax3793, September 2019.

\bibitem{2018PhRvL.120b1101A}
M.~{Aguilar}, L.~{Ali Cavasonza}, G.~{Ambrosi}, et~al.
\newblock {Observation of New Properties of Secondary Cosmic Rays Lithium,
  Beryllium, and Boron by the Alpha Magnetic Spectrometer on the International
  Space Station}.
\newblock {\em Physical Review Letters}, 120(2):021101, January 2018.

\bibitem{2020PhRvL.124u1102A}
M.~{Aguilar}, L.~{Ali Cavasonza}, G.~{Ambrosi}, et~al.
\newblock {Properties of Neon, Magnesium, and Silicon Primary Cosmic Rays
  Results from the Alpha Magnetic Spectrometer}.
\newblock {\em \prl}, 124(21):211102, May 2020.

\bibitem{2021PhRvL.126d1104A}
M.~{Aguilar}, L.~Ali {Cavasonza}, M.~S. {Allen}, et~al.
\newblock {Properties of Iron Primary Cosmic Rays: Results from the Alpha
  Magnetic Spectrometer}.
\newblock {\em \prl}, 126(4):041104, January 2021.

\bibitem{2021PhR...894....1A}
M.~{Aguilar}, L.~{Ali Cavasonza}, G.~{Ambrosi}, et~al.
\newblock {The Alpha Magnetic Spectrometer (AMS) on the international space
  station: Part II - Results from the first seven years}.
\newblock {\em \physrep}, 894:1--116, February 2021.

\bibitem{2012MNRAS.421.1209T}
S.~{Thoudam} and J.~R. {H{\"o}randel}.
\newblock {Nearby supernova remnants and the cosmic ray spectral hardening at
  high energies}.
\newblock {\em \mnras}, 421:1209--1214, April 2012.

\bibitem{2012ApJ...752L..13T}
N.~{Tomassetti}.
\newblock {Origin of the Cosmic-Ray Spectral Hardening}.
\newblock {\em \apjl}, 752:L13, June 2012.

\bibitem{2015PhRvD..91h3012G}
Daniele {Gaggero}, Alfredo {Urbano}, Mauro {Valli}, and Piero {Ullio}.
\newblock {Gamma-ray sky points to radial gradients in cosmic-ray transport}.
\newblock {\em \prd}, 91(8):083012, April 2015.

\bibitem{2016ChPhC..40a5101J}
C.~{Jin}, Y.-Q. {Guo}, and H.-B. {Hu}.
\newblock {Spatial dependent diffusion of cosmic rays and the excess of primary
  electrons derived from high precision measurements by AMS-02}.
\newblock {\em Chinese Physics C}, 40(1):015101, January 2016.

\bibitem{2010ApJ...725..184B}
P.~L. {Biermann}, J.~K. {Becker}, J.~{Dreyer}, et~al.
\newblock {The Origin of Cosmic Rays: Explosions of Massive Stars with Magnetic
  Winds and Their Supernova Mechanism}.
\newblock {\em \apj}, 725:184--187, December 2010.

\bibitem{2014A&A...567A..33T}
S.~{Thoudam} and J.~R. {H{\"o}randel}.
\newblock {GeV-TeV cosmic-ray spectral anomaly as due to reacceleration by weak
  shocks in the Galaxy}.
\newblock {\em \aap}, 567:A33, July 2014.

\bibitem{2019PhRvL.122d1102A}
M.~{Aguilar}, L.~{Ali Cavasonza}, G.~{Ambrosi}, et~al.
\newblock {Towards Understanding the Origin of Cosmic-Ray Positrons}.
\newblock {\em \prl}, 122(4):041102, February 2019.

\bibitem{2012ApJ...750....3A}
M.~{Ackermann}, M.~{Ajello}, W.~B. {Atwood}, et~al.
\newblock {Fermi-LAT Observations of the Diffuse {$\gamma$}-Ray Emission:
  Implications for Cosmic Rays and the Interstellar Medium}.
\newblock {\em \apj}, 750:3, May 2012.

\bibitem{2016ApJ...819...54G}
Yi-Qing {Guo}, Zhen {Tian}, and Chao {Jin}.
\newblock {Spatial-dependent Propagation of Cosmic Rays Results in the Spectrum
  of Proton, Ratios of P/P, and B/C, and Anisotropy of Nuclei}.
\newblock {\em \apj}, 819(1):54, March 2016.

\bibitem{2018PhRvD..97f3008G}
Yi-Qing {Guo} and Qiang {Yuan}.
\newblock {Understanding the spectral hardenings and radial distribution of
  Galactic cosmic rays and Fermi diffuse {\ensuremath{\gamma}} rays with
  spatially-dependent propagation}.
\newblock {\em \prd}, 97(6):063008, March 2018.

\bibitem{2023PhRvL.130u1002A}
M.~{Aguilar}, L.~{Ali Cavasonza}, B.~{Alpat}, et~al.
\newblock {Properties of Cosmic-Ray Sulfur and Determination of the Composition
  of Primary Cosmic-Ray Carbon, Neon, Magnesium, and Sulfur: Ten-Year Results
  from the Alpha Magnetic Spectrometer}.
\newblock {\em \prl}, 130(21):211002, May 2023.

\bibitem{2018PhRvL.121e1103A}
M.~{Aguilar}, L.~{Ali Cavasonza}, B.~{Alpat}, et~al.
\newblock {Precision Measurement of Cosmic-Ray Nitrogen and its Primary and
  Secondary Components with the Alpha Magnetic Spectrometer on the
  International Space Station}.
\newblock {\em \prl}, 121(5):051103, August 2018.

\bibitem{2021PhRvL.127b1101A}
M.~{Aguilar}, L.~Ali {Cavasonza}, B.~{Alpat}, et~al.
\newblock {Properties of a New Group of Cosmic Nuclei: Results from the Alpha
  Magnetic Spectrometer on Sodium, Aluminum, and Nitrogen}.
\newblock {\em \prl}, 127(2):021101, July 2021.

\bibitem{2021PhRvL.126h1102A}
M.~{Aguilar}, L.~Ali {Cavasonza}, M.~S. {Allen}, et~al.
\newblock {Properties of Heavy Secondary Fluorine Cosmic Rays: Results from the
  Alpha Magnetic Spectrometer}.
\newblock {\em \prl}, 126(8):081102, February 2021.

\bibitem{2023RAA....23k5002P}
Xu~{Pan} and Qiang {Yuan}.
\newblock {Injection Spectra of Different Species of Cosmic Rays from AMS-02,
  ACE-CRIS and Voyager-1}.
\newblock {\em Research in Astronomy and Astrophysics}, 23(11):115002, November
  2023.

\bibitem{1964ocr..book.....G}
V.~L. {Ginzburg} and S.~I. {Syrovatskii}.
\newblock {\em {The Origin of Cosmic Rays}}.
\newblock 1964.

\bibitem{1971ApJ...170..265S}
John {Skilling}.
\newblock {Cosmic Rays in the Galaxy: Convection or Diffusion?}
\newblock {\em \apj}, 170:265, December 1971.

\bibitem{berezinskii1990dogiel}
Bulanov Berezinskii.
\newblock Dogiel, et ptuskin. astrophysics of cosmic rays, 1990.

\bibitem{2008JCAP...10..018E}
C.~{Evoli}, D.~{Gaggero}, D.~{Grasso}, and L.~{Maccione}.
\newblock {Cosmic ray nuclei, antiprotons and gamma rays in the galaxy: a new
  diffusion model}.
\newblock {\em \jcap}, 10:018, October 2008.

\bibitem{2017Sci...358..911A}
A.~U. {Abeysekara}, A.~{Albert}, R.~{Alfaro}, et~al.
\newblock {Extended gamma-ray sources around pulsars constrain the origin of
  the positron flux at Earth}.
\newblock {\em Science}, 358:911--914, November 2017.

\bibitem{2021PhRvL.126x1103A}
F.~{Aharonian}, Q.~{An}, L.~X. {Axikegu}, Bai, et~al.
\newblock {Extended Very-High-Energy Gamma-Ray Emission Surrounding PSR J 0622
  +3749 Observed by LHAASO-KM2A}.
\newblock {\em \prl}, 126(24):241103, June 2021.

\bibitem{1998ApJ...504..761C}
Gary~L. {Case} and Dipen {Bhattacharya}.
\newblock {A New {\ensuremath{\Sigma}}-D Relation and Its Application to the
  Galactic Supernova Remnant Distribution}.
\newblock {\em \apj}, 504(2):761--772, September 1998.

\bibitem{1968ApJ...154.1011G}
L.~J. {Gleeson} and W.~I. {Axford}.
\newblock {Solar Modulation of Galactic Cosmic Rays}.
\newblock {\em \apj}, 154:1011, December 1968.

\bibitem{2017PhRvL.119y1101A}
M.~{Aguilar}, L.~{Ali Cavasonza}, B.~{Alpat}, et~al.
\newblock {Observation of the Identical Rigidity Dependence of He, C, and O
  Cosmic Rays at High Rigidities by the Alpha Magnetic Spectrometer on the
  International Space Station}.
\newblock {\em \prl}, 119(25):251101, December 2017.

\bibitem{2019PhRvL.123r1102A}
M.~{Aguilar}, L.~{Ali Cavasonza}, G.~{Ambrosi}, et~al.
\newblock {Properties of Cosmic Helium Isotopes Measured by the Alpha Magnetic
  Spectrometer}.
\newblock {\em \prl}, 123(18):181102, November 2019.

\bibitem{2022PTEP.2022h3C01W}
R.~L. {Workman}, V.~D. {Burkert}, V.~{Crede}, et~al.
\newblock {Review of Particle Physics}.
\newblock {\em Progress of Theoretical and Experimental Physics},
  2022(8):083C01, August 2022.

\bibitem{2016ApJ...831...18C}
A.~C. {Cummings}, E.~C. {Stone}, B.~C. {Heikkila}, et~al.
\newblock {Galactic Cosmic Rays in the Local Interstellar Medium: Voyager 1
  Observations and Model Results}.
\newblock {\em \apj}, 831(1):18, November 2016.

\end{thebibliography}

\end{document}